\documentclass[aps,pra,twocolumn,showpacs]{revtex4}
\usepackage{amssymb}
\usepackage{amsmath}
\usepackage{graphicx}
\usepackage{pstricks}
\usepackage{epsfig}

\begin{document}

%%  AUTHOR'S MACRO
\def\beq{\begin{equation}}
\def\eeq{\end{equation}}
\def\bleq{\begin{eqnarray}}
\def\eleq{\end{eqnarray}} 
\newcommand{\Tr}{{\rm Tr}} 
\newcommand{\tr}{{\rm tr}} 
\newcommand{\sgn}{{\rm sgn}} 
\newcommand{\mean}[1]{\langle #1 \rangle}
\newcommand{\const}{{\rm const}} 
\newcommand{\ie}{i.e. }
\newcommand{\eg}{e.g. }
\newcommand{\cc}{{\rm c.c.}} 
\newcommand{\hc}{{\rm h.c.}} 
\def\eps{\epsilon}
\def\varphibf{\boldsymbol{\varphi}}
\def\half{\frac{1}{2}}

\def\n{{\bf n}} 
\def\p{{\bf p}} 
\def\q{{\bf q}}
\def\r{{\bf r}}
\def\A{{\bf A}}
\def\nablabf{\boldsymbol{\nabla}}
\def\sigmabf{\boldsymbol{\sigma}} 
\def\Pibf{\boldsymbol{\Pi}}
\def\pibf{\boldsymbol{\pi}}

\def\para{\parallel}

\def\w{\omega}
\def\wn{\omega_n}
\def\wnu{\omega_\nu}
\def\dt{\partial_t}
\def\dx{\partial_x}
\def\dy{\partial_y} 
\def\dtau{{\partial_\tau}}  
\def\det{{\rm det}} 
\def\intr{\int d^dr}  
\def\inttau{\int_0^\beta d\tau}
\def\intinf{\int_{-\infty}^\infty}
\def\dintinf{\displaystyle \int_{-\infty}^\infty}
\def\intw{\int_{-\infty}^\infty \frac{d\w}{2\pi}}

\def\calC{{\cal C}}
\def\calO{{\cal O}}

\def\Sign{\Sigma_{\rm n}}
\def\Sigan{\Sigma_{\rm an}}
\def\Gll{G_{\rm ll}}
\def\Gtt{G_{\rm tt}}
\def\Glt{G_{\rm lt}}
\def\Gbarll{\bar G_{\rm ll}}
\def\Gbartt{\bar G_{\rm tt}}
\def\Gbarlt{\bar G_{\rm lt}}
\def\Gtildell{\tilde G_{\rm ll}}
\def\Gtildett{\tilde G_{\rm tt}}
\def\Gtildelt{\tilde G_{\rm lt}}
\def\GllB{G^{\rm B}_{\rm ll}}
\def\GttB{G^{\rm B}_{\rm tt}}
\def\GltB{G^{\rm B}_{\rm lt}}
\def\Ill{I_{\rm ll}}
\def\Itt{I_{\rm tt}}
\def\Ibarll{\bar I_{\rm ll}}
\def\Ibartt{\bar I_{\rm tt}}
\def\Itildell{\tilde I_{\rm ll}}
\def\Itildett{\tilde I_{\rm tt}}
\def\Jllll{J_{\rm ll,ll}}
\def\Jtttt{J_{\rm tt,tt}}
\def\Jltlt{J_{\rm lt,lt}}
\def\Jlltt{J_{\rm ll,tt}}
\def\Jttll{J_{\rm tt,ll}}
\def\Jlllt{J_{\rm ll,lt}}
\def\Jltll{J_{\rm lt,ll}}
\def\Jttlt{J_{\rm tt,lt}}
\def\Jlttt{J_{\rm lt,tt}}
\def\Jbarllll{\bar J_{\rm ll,ll}}
\def\Jbartttt{\bar J_{\rm tt,tt}}
\def\Jbarltlt{\bar J_{\rm lt,lt}}
\def\Jbarlltt{\bar J_{\rm ll,tt}}
\def\Jbarttll{\bar J_{\rm tt,ll}}
\def\Jbarlllt{\bar J_{\rm ll,lt}}
\def\Jbarltll{\bar J_{\rm lt,ll}}
\def\Jbarttlt{\bar J_{\rm tt,lt}}
\def\Jbarlttt{\bar J_{\rm lt,tt}}
\def\Jtildellll{\tilde J_{\rm ll,ll}}
\def\Jtildetttt{\tilde J_{\rm tt,tt}}
\def\Jtildeltlt{\tilde J_{\rm lt,lt}}
\def\Jtildelltt{\tilde J_{\rm ll,tt}}
\def\Jtildettll{\tilde J_{\rm tt,ll}}
\def\Jtildelllt{\tilde J_{\rm ll,lt}}
\def\Jtildeltll{\tilde J_{\rm lt,ll}}
\def\Jtildettlt{\tilde J_{\rm tt,lt}}
\def\Jtildelttt{\tilde J_{\rm lt,tt}}
\def\All{A_{\rm ll}}
\def\Att{A_{\rm tt}}
\def\Alt{A_{\rm lt}}
\def\An{A_{\rm n}}
\def\Aan{A_{\rm an}}
\def\Gn{G_{\rm n}}
\def\Gan{G_{\rm an}}

%%%%%%%%%%%%%%%%%%%%%%%%%%%%%%%%%%%%%%%%%%%%%%%%%%%%%%%%%%%%%%%%%%%%%%%%%%%%%%%%%%%%%

\title{Infrared behavior and spectral function of a Bose superfluid at zero temperature}

\author{N. Dupuis}
\affiliation{
 Laboratoire de Physique Th\'eorique de la Mati\`ere Condens\'ee, 
CNRS UMR 7600, \\ Universit\'e Pierre et Marie Curie, 4 Place Jussieu, 
75252 Paris Cedex 05,  France}
\affiliation{Laboratoire de Physique des Solides, CNRS UMR 8502, Universit\'e Paris-Sud, 91405 Orsay, France}

\date{October 22, 2009}
\begin{abstract} 
In a Bose superfluid, the coupling between transverse (phase) and longitudinal fluctuations leads to a divergence of the longitudinal correlation function, which is responsible for the occurrence of infrared divergences in the perturbation theory and the breakdown of the Bogoliubov approximation. We report a non-perturbative renormalization-group (NPRG) calculation of the one-particle Green function of an interacting boson system at zero temperature. We find two regimes separated by a characteristic momentum scale $k_G$ (``Ginzburg'' scale). While the Bogoliubov approximation is valid at large momenta and energies, $|\p|,|\w|/c\gg k_G$ (with $c$ the velocity of the Bogoliubov sound mode), in the infrared (hydrodynamic) regime $|\p|,|\w|/c\ll k_G$ the normal and anomalous self-energies exhibit singularities reflecting the divergence of the longitudinal correlation function. In particular, we find that the anomalous self-energy agrees with the Bogoliubov result $\Sigan(\p,\w)\simeq\const$ at high-energies and behaves as $\Sigan(\p,\w)\sim (c^2\p^2-\w^2)^{(d-3)/2}$ in the infrared regime (with $d$ the space dimension), in agreement with the Nepomnyashchii identity $\Sigan(0,0)=0$ and the predictions of Popov's hydrodynamic theory. We argue that the hydrodynamic limit of the one-particle Green function is fully determined by the knowledge of the exponent $3-d$ characterizing the divergence of the longitudinal susceptibility and the Ward identities associated to gauge and Galilean invariances. The infrared singularity of $\Sigan(\p,\w)$ leads to a continuum of excitations (coexisting with the sound mode) which shows up in the one-particle spectral function.
\end{abstract}

\maketitle

\section{Introduction}

Following the success of the Bogoliubov theory~\cite{Bogoliubov47} in providing a microscopic explanation of superfluidity, much theoretical work has been devoted to the understanding of the infrared behavior and the calculation of the one-particle Green function of a Bose superfluid~\cite{note4}. Early attempts to improve the Bogoliubov approximation encountered difficulties due to a singular perturbation theory plagued by infrared divergences~\cite{Beliaev58a,Beliaev58b,Hugenholtz59,Gavoret64}. Although these divergences cancel out in local gauge invariant physical quantities (condensate density, Goldstone mode velocity, etc.), they do have a definite physical origin: they reflect the divergence of the longitudinal susceptibility which is induced by the coupling between transverse (phase) and longitudinal fluctuations. This is a general phenomenon~\cite{Patasinskij73} in systems with a continuous broken symmetry as discussed at the end of this section. 

Using field-theoretical diagrammatic methods to handle the infrared divergences of the perturbation theory, Nepomnyashchii and Nepomnyashchii (NN) showed that one of the fundamental quantities of a Bose superfluid, the anomalous self-energy $\Sigan(p)$, vanishes in the limit $p=(\p,\w)\to 0$ in dimension $d\leq 3$,  even though the low-energy mode remains phonon-like with a linear spectrum~\cite{Nepomnyashchii75,Nepomnyashchii78,Nepomnyashchii83}. This exact result shows that the Bogoliubov approximation, where the linear spectrum and the superfluidity rely on a finite value of the anomalous self-energy, breaks down at low energy. 

An alternative approach to superfluidity, based on a phase-amplitude representation of the boson field, has been proposed by Popov~\cite{Popov_book_2}. This approach is free of infrared singularity, but restricted to the (low-momentum) hydrodynamic regime and therefore does not allow to study the high-momentum or high-frequency regime where the Bogoliubov approximation is expected to be valid. Nevertheless, Popov's theory~\cite{Popov79} agrees with the asymptotic low-energy behavior of $\Sigan(p)$ obtained by NN~\cite{Nepomnyashchii75,Nepomnyashchii78,Nepomnyashchii83}. Furthermore, the expression of the anomalous self-energy obtained by NN and Popov in the low-energy limit yields a continuum of (one-particle) excitations coexisting with the Bogoliubov sound mode~\cite{Giorgini92}, in marked contrast with the Bogoliubov theory where the sound mode is the sole excitation at low energy. 

The instability of the Bogoliubov fixed point in dimension $d\leq 3$ towards a different fixed point characterized by the divergence of the longitudinal susceptibility has been confirmed by Castellani {\it et al.}~\cite{Castellani97,Pistolesi04}. Using the Ward identities associated to the local gauge symmetry and a renormalization group approach, these authors obtained the exact infrared behavior of a Bose superfluid at zero temperature. Related results, both at zero~\cite{Wetterich08,Dupuis07,Sinner09,Dupuis09} and finite~\cite{Andersen99,Floerchinger08,Floerchinger09a,Floerchinger09b,Eichler09} temperature have been obtained by several authors within the framework of the non-perturbative renormalization group.

In this paper, we study a weakly interacting Bose superfluid at zero temperature using the so-called BMW NPRG scheme introduced by Blaizot, M\'endez-Galain and Wschebor~\cite{Blaizot06,Benitez09}. Compared to more traditional RG approaches, the NPRG approach presents a number of advantages: i) symmetries are naturally implemented (by a proper Ansatz for the effective action or the two-point vertex (Sec.~\ref{sec_ea})) and Ward identities are naturally satisfied; ii) the NPRG approach is not restricted to the low-energy asymptotic behavior but can deal with all energy scales. In particular, it relates physical quantities at a macroscopic scale to the parameters of the microscopic model; iii) the BMW scheme enables to obtain the full momentum and frequency dependence of the correlation functions~\cite{Dupuis09}. 

In the NPRG approach, the main quantities of interest are the average effective action $\Gamma_k$ (the generating functional of one-particle irreducible vertices) and its second-order functional derivative, the two-point vertex $\Gamma_k^{(2)}(p)$, whose inverse gives the one-particle propagator (Sec.~\ref{sec_ea}). Fluctuations beyond the Bogoliubov approximation are gradually taken into account as the (running) momentum scale $k$ is lowered from the microscopic scale $\Lambda$ down to zero. In Sec.~\ref{sec_de}, we show that the infrared behavior of the one-particle propagator is entirely determined by the Ward identities associated to the Galilean and local gauge invariances of the microscopic action, and the exponent $3-d$ characterizing the divergence of the longitudinal susceptibility (see the discussion at the end of the Introduction). In Sec.~\ref{sec_rg} we derive the BMW flow equations satisfied by the two-point vertex $\Gamma^{(2)}(p)$ and obtain the analytical solution in the infrared regime. Numerical results are discussed in Sec.~\ref{sec_nr}. We find that the Bogoliubov approximation breaks down at a characteristic momentum length $k_G$ (``Ginzburg'' scale) which, for weak boson-boson interactions, is much smaller than the inverse healing length $k_h$ ($k_h$ is defined in Appendix \ref{app_bog}). Although local gauge invariant quantities are not sensitive to $k_G$, the effective action $\Gamma_k$ is attracted to a fixed point characterized by the divergence of the longitudinal susceptibility when $k\ll k_G$. We discuss in detail the frequency and momentum dependence of the two-point vertex $\Gamma^{(2)}_k(p)$. While for $|\p|\gg k_G$ or $|\w|/c\gg k_G$ (with $c$ the velocity of the Bogoliubov sound mode), $\Gamma^{(2)}_k(p)$ is well described by the Bogoliubov approximation, we reproduce the low-energy asymptotic behavior obtained by NN when $|\p|,|\w|/c\ll k_G$. In this regime, the longitudinal correlation function becomes singular and its spectral function exhibits a continuum of one-particle excitations in agreement with the predictions of Popov's hydrodynamic theory. Thus our approach provides a unified picture of superfluidity in interacting boson systems and connects Bogoliubov's theory to Popov's hydrodynamic theory. In the conclusion, we comment about a possible extension of our results to strongly interacting or one-dimensional superfluids.

\section*{Infrared behavior of interacting bosons} 

Since the divergence of the longitudinal susceptibility plays a key role in the infrared behavior of interacting boson systems, we first discuss this phenomenon both in classical and quantum systems (for a pedagogical discussion, see also Ref.~\cite{Weichman88}). Let us first consider a $\varphibf^4$ theory defined by the action 
\beq
S = \half \int d^dr \left\lbrace (\nabla\varphibf)^2 + v \varphibf^2 + \frac{\lambda}{4}\bigr(\varphibf^2\bigr)^2 \right\rbrace ,
\label{action1}
\eeq
where $\varphibf(\r)$ is a real $N$-component field and $d$ the space dimension. When $v<0$, the mean-field (saddle-point) analysis predicts a non-zero order parameter $\varphibf_0=\mean{\varphibf}$. Including Gaussian fluctuations about the saddle-point $\varphibf_0$, we find a gapped mode and $N-1$ Goldstone modes corresponding to longitudinal ($\delta\varphibf \parallel \varphibf_0$) and transverse ($\delta\varphibf \perp \varphibf_0$) fluctuations, respectively. The correlation functions read
\beq
\begin{split}
G_\parallel(\p) &= \frac{1}{\p^2+\lambda\varphibf_0^2} , \\ 
G_\perp(\p) &= \frac{1}{\p^2} .
\end{split}
\label{cor1}
\eeq
This result, which neglects interactions between longitudinal and transverse fluctuations, is incorrect. In the ordered phase, the amplitude fluctuations of $\varphibf$ are gapped and the low-energy effective description is a non-linear $\sigma$ model~\cite{Zinn_book} 
\beq
S[\n] = \frac{\rho}{2} \int d^dr (\nablabf \n)^2 ,
\label{action2}
\eeq
where $\n$ is a unit vector ($\n^2=1$). To a first approximation, equation (\ref{action2}) can be obtained by setting $\varphibf = |\varphibf_0| \n$ in (\ref{action1}) (which gives $\rho=\varphibf^2_0$). The non-linear $\sigma$ model is solved by writing $\n=(\sigma,\pibf)$ in terms of its longitudinal and transverse components ($\n\cdot\varphibf_0=\sigma|\varphibf_0|$ and $\pibf\perp\varphibf_0$). In the low-energy limit, the action (\ref{action2}) describes $N-1$ non-interacting Goldstone modes with propagator $G_\perp(\p)\sim 1/\p^2$. The longitudinal propagator can be obtained from the constraint $\n^2=\sigma^2+\pibf^2=1$, i.e. $\sigma\simeq 1-\frac{\pibf^2}{2}$, 
\beq
G_\para(\r) = \mean{\sigma(\r)\sigma(0)}_c \approx \frac{N-1}{2} G^2_\perp(\r) \sim \frac{1}{|\r|^{2d-4}} ,
\label{cor2}
\eeq
where $\mean{\cdots}_c$ stands for the connected part of the propagator and $G_\perp(\r)\sim 1/|\r|^{d-2}$ denotes the transverse propagator in real space. Equation (\ref{cor2}) is obtained by using Wick's theorem. In Fourier space, we thus obtain
\beq
G_\para(\p) \sim \frac{1}{|\p|^{4-d}} \quad (\p\to 0) 
\label{cor3}
\eeq
for $d<4$ and a logarithmic divergence for $d=4$. Contrary to the predictions of Gaussian theory [Eq.~(\ref{cor1})], the longitudinal susceptibility is not finite but diverges for $\p\to 0$ when $d\leq 4$~\cite{Patasinskij73,Fisher73,Anishetty99}. This divergence is weaker than that of the transverse propagator for all dimensions larger than the lower critical dimension $d_L=2$. The appearance of a singularity in the longitudinal channel, driven by transverse fluctuations, is a general phenomenon in systems with a continuous broken symmetry~\cite{Patasinskij73}. The momentum scale $k_G$ below which the Gaussian approximation (\ref{cor1}) breaks down is exponentially small for $d=4$ (and $\lambda\to 0$) and of order $\lambda^{1/(4-d)}$ for $d<4$ (see Appendix \ref{subsec_kG} for the estimation of $k_G$ in a Bose superfluid).

The same conclusion can be drawn from the NPRG analysis of the ordered phase of the linear model (\ref{action1}). The NPRG predicts the coupling constant to scale as $\lambda\sim k^{4-d}$ where $k$ is a running momentum scale~\cite{Berges02}. This scaling follows from the flowing of the dimensionless coupling constant $\tilde\lambda=\lambda k^{d-4}$ to a finite value $\tilde\lambda^*$ for $k\to 0$ ($\tilde\lambda\sim 1/\ln k$ for $d=4$). The longitudinal propagator then diverges as $G_\para(\p=0)\sim 1/\lambda\varphibf_0^2\sim 1/k^{4-d}$ and, identifying $k$ with $|\p|$ to extract the $\p$ dependence of the propagator, we reproduce (\ref{cor3}). Thus, the divergence (\ref{cor3}) of the longitudinal susceptibility is a consequence of the fixed point structure of the RG flow in the ordered phase of the linear model (\ref{action1}). 

These considerations easily generalize to a quantum model with the Euclidean action 
\beq
S = \half \inttau \int d^dr \biggl\lbrace  (\nabla\varphibf)^2 + c^{-2}(\dtau \varphibf)^2 + v \varphibf^2 + \frac{\lambda}{4}\bigr(\varphibf^2\bigr)^2 \biggr\rbrace ,
\label{action3}
\eeq
where $\tau$ is an imaginary time varying between 0 and the inverse temperature $\beta=1/T$, and $c$ the velocity of the Goldstone mode. At zero temperature, we expect the (Euclidean) propagator to behave as 
\beq
\begin{split}
G_\perp(\p,i\w) &\sim \frac{1}{\w^2+c^2\p^2} , \\
G_\para(\p,i\w) &\sim \frac{1}{(\w^2+c^2\p^2)^{(3-d)/2}}  ,
\end{split}
\quad (\p,\w\to 0) 
\label{cor4}
\eeq
where $\w$ is a Matsubara frequency. (The divergence of $G_\para$ is logarithmic in three dimensions.) The expression of $G_\para$ follows from (\ref{cor3}) with $|\p|$ replaced by $(\w^2+c^2\p^2)^{1/2}$ and $d$ by the effective dimensionality $d+1$ to account for the imaginary time dependence of the field. As in the classical model (\ref{action1}), it can be justified either from an effective low-energy description based on the (quantum) non-linear $\sigma$ model or directly from the linear model (\ref{action3}). After analytical continuation $i\w\to \w$, the transverse propagator $G_\perp$ exhibits a pole at $\w=\pm c|\p|$. On the contrary, $G_\para$ has no pole-like structure but a branch cut which yields a critical continuum of excitations lying above the Goldstone mode energy $\w=c|\p|$. This continuum results from the decay of a normally massive amplitude mode with momentum $\p$ into a pair of transverse excitations with momenta $\q$ and $\p-\q$~\cite{Zwerger04}. 

Interacting bosons are described by a complex field $\psi$ or, equivalently, a two-component real field $(\psi_1,\psi_2)$. In the ordered phase, the global U(1) symmetry~\cite{note2} is broken, giving rise to a gapless (Goldstone) phase mode (the Bogoliubov sound mode). Although the action of non-relativistic bosons differs from the relativistic-type action (\ref{action3}) (see Eq.~(\ref{action4}) below), the preceding conclusions regarding the longitudinal propagator still hold in the superfluid phase. The reason is that the argument leading to (\ref{cor4}) relies on the existence of a Goldstone mode with linear dispersion $\w=c|\p|$ rather than the precise form of the microscopic action. The one-particle propagator in the superfluid phase is usually expressed in terms of a ``normal'' self-energy, $\Sigma_{\rm n}(\p,i\w)$, and an ``anomalous'' one, $\Sigma_{\rm an}(\p,i\w)$~\cite{AGD_book,Fetter_book}. In Appendix \ref{app_bog}, we show on general grounds that
\beq
G_\para(\p,i\w) \simeq -\frac{1}{2\Sigma_{\rm an}(\p,i\w)} \quad (\p,\w\to 0) .
\eeq
Comparing with (\ref{cor4}), we conclude that the anomalous self-energy 
\beq
\Sigan(\p,i\w) \sim (\w^2+c^2\p^2)^{(3-d)/2} \quad (\p,\w\to 0) 
\label{sig1}
\eeq
is singular at low-energy for $d\leq 3$ (the singularity is logarithmic when $d=3$). This singularity, which also shows up in the normal self-energy, is related to the infrared divergences that were encountered early on in the perturbation theory of interacting boson systems~\cite{Beliaev58a,Beliaev58b,Hugenholtz59,Gavoret64}. The exact result $\Sigan(0,0)=0$ and the asymptotic expression (\ref{sig1}) were first obtained by NN from a field-theoretical (diagrammatic) approach~\cite{Nepomnyashchii75,Nepomnyashchii78,Nepomnyashchii83}. NN's analysis shows that the infrared behavior markedly differs from the predictions of the Bogoliubov theory ($\Sigan(\p,i\w)=\const$). One can estimate the momentum ``Ginzburg'' scale $k_G$ below which the Bogoliubov approximation breaks down from perturbation theory~\cite{Pistolesi04,note8} (see Appendix \ref{app_bog}), 
\beq
k_G \sim \left\lbrace 
\begin{array}{lc}
(gm k_h)^{1/(3-d)} & (d<3) , \\
k_h \exp\left( - \frac{4\sqrt{2}\pi^2}{gmk_h}\right) & (d=3) .
\end{array}
\right.
\label{kg_est}
\eeq
In three dimensions, $k_G$ vanishes exponentially when the dimensionless interaction constant $gmk_h\to 0$. In two dimensions, the vanishing of $k_G$ with $gm$ is only linear. 

It was realized by Popov that a phase-density representation of the boson field $\psi=\sqrt{n}e^{i\theta}$ leads to a theory free of infrared divergences~\cite{Popov_book_2,Nepomnyashchii83}. Popov's theory is based on an hydrodynamic action $S[n,\theta]$ and is valid below a characteristic momentum $k_0$. Since the long-distance physics is governed by the Goldstone (phase) mode, a minimal hydrodynamic description would start from the phase-only action
\beq
S = \frac{n_s}{2m} \inttau \intr \left[(\nablabf\theta)^2 + c^{-2} (\dtau\theta)^2 \right] , 
\label{action_the} 
\eeq
where $n_s$ is the superfluid density. This action can be obtained from the hydrodynamic action $S[n,\theta]$ by integrating out the density field. It is equivalent to that of the non-linear $\sigma$ model in the $O(N)$ model [Eq.~(\ref{action2})]. Writing $\psi\simeq \sqrt{n}e^{i\theta}$ and expanding with respect to phase fluctuations (with the boson density $n=\const$), one finds
\beq
\begin{split}
G_\perp(\r\tau) &= n G_{\theta\theta}(\r\tau) , \\
G_\para(\r\tau) &= \frac{n}{2} G_{\theta\theta}(\r\tau)^2 
\end{split}
\label{cor5}
\eeq
for the propagator of the transverse ($\delta\psi=i\sqrt{n}\theta$) and longitudinal ($\delta\psi=-\sqrt{n}\theta^2/2$) fluctuations, respectively. $G_{\theta\theta}$ is the phase propagator whose Fourier transform $(m/n_s)(\p^2+\w^2/c^2)^{-1}$ is read off from (\ref{action_the}). In Fourier space, equations (\ref{cor5}) coincide with (\ref{cor4}). Thus Popov's approach reproduces the infrared behavior (\ref{sig1}) obtained by NN~\cite{Popov79}. The determination of the characteristic momentum $k_0$ below which the hydrodynamic approach is valid is non-trivial in the Popov approach as it requires to integrate out all modes with momenta $|\p|>k_0$ to obtain the low-energy hydrodynamic description~\cite{Popov_book_2,Popov79}.  Interestingly, $k_0$ coincides with the Ginzburg scale $k_G$~\cite{Pistolesi04}.

\section{The average effective action}
\label{sec_ea}

We consider interacting bosons at zero temperature, with the action
\beq
S = \int dx \left[ \psi^*(x)\left(\dtau-\mu - \frac{\nablabf^2}{2m}
  \right) \psi(x) + \frac{g}{2} |\psi(x)|^4 \right]
\label{action4}
\eeq
($\hbar=k_B=1$ throughout the paper), where $\psi(x)$ is a bosonic (complex) field, $x=(\r,\tau)$, and $\int dx=\inttau \int d^dr$. $\tau\in [0,\beta]$ is an imaginary time, $\beta\to\infty$ the inverse temperature, and $\mu$ denotes the
chemical potential. The interaction is assumed to be local in space and the
model is regularized by a momentum cutoff $\Lambda$. We assume the coupling constant $g$ to be weak (dimensionless coupling constant $gm \bar n^{1-2/d}\ll 1$, with $\bar n$ the mean density) and consider a space dimension $d$ larger than 1. It will often be convenient to write the boson field
\beq
\psi(x) = \frac{1}{\sqrt{2}} [\psi_1(x) + i\psi_2(x)]
\label{psidef}
\eeq
in terms of two real fields $\psi_1$ and $\psi_2$. 

To define the average effective action~\cite{Berges02}, we add to the action (\ref{action4}) a source term $-\int dx(J^*\psi+\cc)$ and an infrared regulator 
\beq
\Delta S_k = \int dxdx' \psi^*(x) R_k(x-x') \psi(x') 
\eeq
which suppresses fluctuations with momenta and energies below a characteristic scale $k$ but leaves the high-momenta/frequencies modes unaffected. The average effective action 
\begin{align}
\Gamma_k[\phi^*,\phi] ={}& - \ln Z_k[J^*,J] + \int dx [J^*(x)\phi(x) + \cc] \nonumber \\ &- \Delta S_k[\phi^*,\phi].
\label{eadef}
\end{align} 
is defined as the Legendre transform of $-\ln Z_k[J^*,J]$, 
($Z_k[J^*,J]$ is the partition function) minus the regulator term $\Delta S_k[\phi^*,\phi]$. Here $\phi^{(*)}(x)=\mean{\psi^{(*)}(x)}_{J^*,J}$ is the superfluid order parameter.

The effective action (\ref{eadef}) is the generating functional of the one-particle irreducible vertices. The infrared regulator $R_k$ is chosen such that for $k=\Lambda$ all fluctuations are frozen. The mean-field theory, where the effective action $\Gamma_\Lambda[\phi^*,\phi]$ reduces to the microscopic action $S[\phi^*,\phi]$, becomes exact thus reproducing the result of the Bogoliubov approximation [See Eqs.~(\ref{flowinit}) below]. On the other hand for $k=0$, provided that $R_{k=0}$ vanishes, $\Gamma_k$ gives the effective action of the original model (\ref{action4}) and allows us to obtain all physical quantities of interest. In practice we take the regulator 
\beq
R_k(p) = \frac{Z_{A,k}}{2m} \left(\p^2+\frac{\w^2}{c_0^2} \right) r\left( \frac{\p^2}{k^2}+\frac{\w^2}{k^2c_0^2} \right) , 
\label{regdef}
\eeq
where $r(Y)=(e^Y-1)^{-1}$ and $p=(\p,i\w)$. The $k$-dependent variable $Z_{A,k}$ is defined below. A natural choice for the velocity $c_0$ would be the actual ($k$-dependent) velocity of the Goldstone mode. In the weak coupling limit, however, the Goldstone mode velocity renormalizes only weakly and is well approximated by the $k$-independent value $c_0=\sqrt{g\bar n/m}$ ($\bar n$ is the mean boson density). The regulator (\ref{regdef}) differs from previous works where $R_k(p)$ was taken frequency independent~\cite{Wetterich08,Dupuis07,Sinner09,Dupuis09}. The motivation for the choice (\ref{regdef}) will appear clearly when we will discuss the BMW NPRG scheme.  

We are primarily interested in the effective potential
\beq
U(\phi) = \frac{1}{\beta V} \Gamma[\phi] \Bigl|_{\phi\,\rm const} 
\eeq
($V$ is the volume of the system) and the two-point vertex 
\beq
\Gamma^{(2)}_{ij}(x,x';\phi) = \frac{\delta^{(2)}\Gamma[\phi]}{\delta\phi_i(x) \delta\phi_j(x')}\biggl|_{\phi\,\rm const} 
\eeq
computed in a constant, i.e. uniform and time-independent, field. To alleviate the notations, we now drop the $k$ index. We consider $\phi=(\phi_1,\phi_2)$ as a two-component real field [see Eq.~(\ref{psidef})]. $U$ and $\Gamma^{(2)}_{ij}$ are strongly constrained by the global U(1) invariance of the microscopic action (\ref{action4})~\cite{note2}. The effective potential $U(n)$ must be invariant in this transformation and is therefore a function of the condensate density $n=\half(\phi_1^2+\phi_2^2)$. The actual ($k$-dependent) condensate density $n_0$ is obtained by minimizing the effective potential
\beq
U'(n_0) = 0 .
\label{physstat}
\eeq
Equation (\ref{physstat}) defines the equilibrium state of the system. On the other hand, $\Gamma^{(2)}_{ij}$ must transform as a tensor when the two-dimensional vector $(\phi_1,\phi_2)$ is rotated by an arbitrary angle $\alpha$. Since one can form three tensors, $\delta_{i,j}$, $\eps_{i,j}$ and $\phi_i\phi_j$, from the two-dimensional vector $(\phi_1,\phi_2)$, the most general form of the two-point vertex is~\cite{note1}
\beq
\Gamma^{(2)}_{ij}(p;\phi) = \delta_{i,j} \Gamma_A(p;n) + \phi_i\phi_j \Gamma_B(p;n) + \eps_{ij} \Gamma_C(p;n) 
\label{sym1}
\eeq
in Fourier space. $\eps_{ij}$ denotes the antisymmetric tensor. In addition, parity and time reversal invariance implies
\beq
\begin{split}
\Gamma^{(2)}_{ij}(\p,i\w;\phi) &= \Gamma^{(2)}_{ij}(-\p,i\w;\phi) , \\ 
\Gamma^{(2)}_{ij}(\p,i\w;\phi) &= (2\delta_{i,j}-1)\Gamma^{(2)}_{ij}(\p,-i\w;\phi^*) ,
\end{split}
\label{sym2}
\eeq
where $\phi=(\phi_1,\phi_2)$ and $\phi^*=(\phi_1,-\phi_2)$. From (\ref{sym1}) and (\ref{sym2}) we deduce 
\beq
\begin{split}
\Gamma_{A}(p;n) &= \Gamma_{A}(-p;n) = \Gamma_{A}(\p,-i\w;n), \\ 
\Gamma_{B}(p;n) &= \Gamma_{B}(-p;n) = \Gamma_{B}(\p,-i\w;n), \\ 
\Gamma_{C}(p;n) &= -\Gamma_{C}(-p;n) = -\Gamma_{C}(\p,-i\w;n) .
\end{split}
\label{sym} 
\eeq 
For $p=0$, we can relate the two-point vertex to the derivatives of the effective potential,
\beq
\Gamma^{(2)}_{ij}(p=0;\phi) = \frac{\partial^2 U(n)}{\partial\phi_i\partial\phi_j} = \delta_{i,j}U'(n) +\phi_i\phi_j U''(n),
\eeq
so that 
\beq
\begin{split}
\Gamma_{A}(p=0;n) &= U'(n) , \\ 
\Gamma_{B}(p=0;n) &= U''(n) , \\
\Gamma_{C}(p=0;n) &= 0 .
\end{split}
\eeq
For $k=\Lambda$, one has $\Gamma_k[\phi]=S[\phi]$ and therefore 
\beq
\begin{split}
U_{k=\Lambda}(n) &= -\mu n + \frac{g}{2} n^2 = - \half gn_0^2 + \frac{g}{2} (n-n_0)^2 , \\ 
\Gamma_{A,k=\Lambda}(p;n) &= \eps_\p + g(n-n_0), \\
\Gamma_{B,k=\Lambda}(p;n) &= g , \\ 
\Gamma_{C,k=\Lambda}(p;n) &= \w , 
\end{split} 
\label{flowinit} 
\eeq
where $n_0\equiv n_{0,k=\Lambda}=\mu/g$. 

We can also relate the two-point vertex to the normal and anomalous self-energies that are usually introduced in the theory of superfluidity~\cite{AGD_book,Fetter_book},  
\beq
\begin{split}
\Sigma_{\rm n}(p) ={}& i\w+\mu-\eps_\p +\bar\Gamma_{A}(p)  +n_{0}\bar\Gamma_{B}(p)-i\bar\Gamma_{C}(p) , \\ 
\Sigma_{\rm an}(p) ={}& n_{0} \bar\Gamma_{B}(p) 
\end{split}
\label{sigrel}
\eeq
(see Appendix \ref{app_bog}),  where $\bar\Gamma_{\alpha}(p)=\Gamma_{\alpha}(p;n_0)$ ($\alpha=A,B,C$) denotes the two-point vertex in the equilibrium state ($n=n_{0}$). 

In the equilibrium state ($n=n_{0}$), the transverse part~\cite{note1} $\bar\Gamma_{A}(p=0)=U'(n_0)$ of the two-point vertex vanishes. This result is a consequence of the invariance of the effective action $\Gamma[\phi]$ in a global U(1) transformation and reflects the existence of a (gapless) Goldstone mode. When expressed in terms of the normal and anomalous self-energies (\ref{sigrel}) (with the condition $\bar\Gamma_C(p=0)=0$), it reproduces the Hugenholtz-Pines theorem~\cite{Hugenholtz59} (\ref{hpth}).

\section{Derivative expansion and infrared behavior} 
\label{sec_de}

On the basis of the arguments given in the Introduction, we expect the anomalous self-energy
\beq
\bar\Gamma_{k=0,B}(p) \sim (\w^2+c^2\p^2)^{(3-d)/2} \quad (\p,\w\to 0) 
\label{sing1}
\eeq
to be singular in the low-energy limit (see Eqs.~(\ref{sig1}) and (\ref{sigrel})). From (\ref{sing1}), we infer 
\beq
\bar\Gamma_{B}(p=0) = U''(n_0) \sim k^{3-d} \quad (k\to 0) .
\label{sing2}
\eeq
Equation (\ref{sing2}) will be obtained in Sec.~\ref{sec_rg} from the NPRG equations. In this section, we show that it is sufficient, when combined with Ward identities associated to Galilean and gauge invariances~\cite{Gavoret64,Huang64,Pistolesi04}, to obtain the infrared behavior of the propagators. 

The infrared regulator (\ref{regdef}) ensures that the vertices are regular functions of $p$ for $|\p|\ll k$ and $|\w|/c\ll k$, even when they become singular functions of $(\p,i\w)$ at $k=0$ ($c\simeq c_0$ is the velocity of the Goldstone mode defined below)~\cite{note7}. In the low-energy limit $|\p|,|\w|/c\ll k$, we can then use the derivative expansion
\beq
\begin{split}
\bar\Gamma_{A}(p) &\simeq Z_A \eps_\p + V_A \w^2 , \\
\bar\Gamma_{B}(p) &\simeq U''(n_0) = \lambda , \\
\bar\Gamma_{C}(p) &\simeq Z_C \w ,
\end{split}
\label{expansion}
\eeq 
consistent with $\bar\Gamma_{A}(p=0)=0$ and the symmetry properties (\ref{sym}). To obtain (\ref{expansion}) we have expanded $\bar\Gamma_{B}(p)$ only to leading order, dropping the next-order term $Z_B\eps_\p+V_B\w^2$. Because of the singularity (\ref{sing1}), the coefficients $Z_B$ and $V_B$ would diverge for $k\to 0$ contrary to $Z_A$, $Z_C$ and $V_A$ that reach finite values. The justification for neglecting the $p$ dependence of the vertex $\bar\Gamma_{B}$ comes from the fact that for $d>1$, $\lambda=\calO(k^{3-d})$ is a very large energy scale with respect to $\bar\Gamma_A,\bar\Gamma^2_C$ for typical momentum and frequency $|\p|,|\w|/c\sim k$. The $p$ dependence of $\bar\Gamma_B(p)$ does not change the leading behavior of $\bar\Gamma_B(p)\sim \calO(k^{3-d})$ which essentially acts as a large mass term in the propagators. 

\subsection{Goldstone mode velocity and superfluid density}

The excitation spectrum can be obtained from the zeros of the determinant of the $2\times 2$ matrix $\bar\Gamma^{(2)}_{ij}(p)$ (after analytical continuation $i\w\to \w+i0^+$),
\begin{align}
\det\,\bar\Gamma^{(2)}(p) &= \bar\Gamma_{A}(p)[\bar\Gamma_{A}(p) + 2n_{0}\bar\Gamma_{B}(p)] + \bar\Gamma^2_{C}(p) \nonumber \\ 
&\simeq 2n_{0}\bar\Gamma_{B}(0) \bar\Gamma_{A}(p) + \bar\Gamma^2_{C}(p) \nonumber \\ & \simeq 2n_0\lambda (Z_A\eps_\p + V_A\w^2)+(Z_C\w)^2 , 
\label{detgam}
\end{align}
where the approximate equality is obtained using $\bar\Gamma_{B}(p)\sim k^{3-d}$, $\bar\Gamma_{A}(p),\bar\Gamma^2_{C}(p)\sim \p^2,\w^2$ and $|\p|,|\w|/c\ll k$. Equation (\ref{detgam}) agrees with the existence of a Goldstone mode (the Bogoliubov sound mode) with velocity
\beq
c = \left( \frac{Z_A/(2m)}{V_A+Z_C^2/(2\lambda n_0)} \right)^{1/2} .
\label{cdef} 
\eeq

The low-energy expansion (\ref{expansion}) can also be used to define the superfluid density $n_s$. Suppose the phase $\theta(\r)$ of the order parameter $\phi(\r)=\sqrt{2n_0}(\cos\theta(\r),\sin\theta(\r))$ varies slowly in space. To lowest order in $\nablabf\theta$, the average effective action will increase by
\begin{align}
\delta \Gamma &= \half \sum_p \bar\Gamma^{(2)}_{A}(p) \phi_2(-p)\phi_2(p) \nonumber \\ &= n_0 \sum_\p \bar\Gamma^{(2)}_{A}(\p,\w=0) \theta(-\p)\theta(\p) \nonumber \\ 
&= \frac{Z_An_{0}}{2m} \beta \intr (\nablabf\theta)^2 .
\label{stiffness}
\end{align}
Identifying the phase stiffness with the superfluid density~\cite{Chaikin_book}, we obtain
\beq
n_s = Z_A n_0 .
\eeq

\subsection{Symmetries and thermodynamic relations} 

The two-point vertex satisfies the following relations,
\beq
\begin{split}
\frac{\partial}{\partial \p^2} \bar\Gamma_A(p) \Bigl|_{p=0} &= \frac{\bar n}{2m n_0}, \\ 
\frac{\partial}{\partial\w^2} \bar\Gamma_A(p) \Bigl|_{p=0} &= - \frac{1}{2n_0} \frac{\partial^2 U}{\partial\mu^2}\biggl|_{n_0} , \\ 
\frac{\partial}{\partial\w} \bar\Gamma_C(p) \Bigl|_{p=0} &= - \frac{\partial^2 U}{\partial n\partial\mu}\biggl|_{n_0} , 
\end{split}
\label{wi} 
\eeq
which follow from Ward identities associated with Galilean (for the first one) and local gauge (for the last two) invariance (see Appendix \ref{app_wi}). Here we consider the effective potential $U(n,\mu)$ as a function of the two independent variables $n$ and $\mu$. The condensate density $n_0=n_0(\mu)$ is then defined by
\beq
\frac{\partial U}{\partial n}\biggl|_{n_0} = 0 ,
\label{n0_def} 
\eeq
while the mean boson density is obtained from 
\begin{align}
\bar n &= - \frac{d}{d\mu} U(n_0,\mu) \nonumber \\ &= - \frac{\partial U}{\partial \mu}\biggl|_{n_0} - \frac{\partial U}{\partial n}\biggl|_{n_0} \frac{dn_0}{d\mu} = - \frac{\partial U}{\partial \mu}\biggl|_{n_0} ,
\label{nbar} 
\end{align} 
where $d/d\mu$ is a total derivative. Equation (\ref{n0_def}) being valid for any $\mu$, one deduces
\beq
\frac{d}{d\mu} \frac{\partial U}{\partial n}\biggl|_{n_0} = \frac{\partial^2 U}{\partial n\partial\mu}\biggl|_{n_0} + \frac{\partial^2 U}{\partial n^2}\biggl|_{n_0} \frac{dn_0}{d\mu} = 0 . 
\label{n0_def_der}
\eeq
From (\ref{wi}) and (\ref{n0_def_der}), one deduces 
\beq
\begin{split}
n_s &= Z_A n_0 = \bar n , \\
V_A &= - \frac{1}{2n_0} \frac{\partial^2 U}{\partial\mu^2}\biggl|_{n_0} , \\ 
Z_C &= - \frac{\partial^2 U}{\partial n\partial\mu}\biggl|_{n_0} = \lambda \frac{dn_0}{d\mu} .
\end{split}
\label{ward1} 
\eeq
The first of these equations states that in a Galilean invariant superfluid at zero temperature, the superfluid density is given by the full density of the fluid~\cite{Gavoret64}. The velocity (\ref{cdef}) can be rewritten as
\beq
c^2 = \frac{\bar n}{m} \frac{1}{-\frac{\partial^2 U}{\partial\mu^2}\Bigl|_{n_0} + \frac{\partial^2 U}{\partial n^2}\Bigl|_{n_0} \bigl(\frac{dn_0}{d\mu}\bigr)^2}.
\eeq
Comparing with 
\beq
\frac{d\bar n}{d\mu} =  -\frac{\partial^2 U}{\partial\mu^2}\biggl|_{n_0} + \frac{\partial^2 U}{\partial n^2}\biggl|_{n_0} \left(\frac{dn_0}{d\mu}\right)^2 , 
\eeq
we deduce that the Goldstone mode velocity
\beq
c= \left( \frac{\bar n}{m(d\bar n/d\mu)}\right)^{1/2} 
\eeq
is equal to the macroscopic sound velocity~\cite{Gavoret64}.

Since thermodynamic quantities, including the condensate ``compressibility'' $dn_0/d\mu$ should be finite, we deduce from (\ref{ward1}) that $Z_C\sim k^{3-d}$ vanishes in the infrared limit, and 
\beq
\lim_{k\to 0} c = \lim_{k\to 0} \left( \frac{Z_A}{2mV_A} \right)^{1/2} .
\label{velir}
\eeq
Both $Z_A=\bar n/n_0$ and the macroscopic sound velocity $c$ being finite, $V_A$ (which vanishes in the Bogoliubov approximation) takes a non-zero value. In the infrared limit, the $\w^2$ term of $\bar\Gamma_A(p)$ is therefore crucial to maintain a linear spectrum and superfluidity. As discussed in more detail in Sec.~\ref{subsec_analytic}, the expression (\ref{velir}) is a manifestation of the relativistic invariance of the effective action which emerges in the low-energy limit. 

\subsection{One-particle propagator}
\label{subsec_1PP} 

We are now in a position to deduce the infrared limit of the one-particle propagator $G_{ij}$. For symmetry reasons (see Sec.~\ref{sec_ea}), 
\begin{align}
G_{ij}(p;\phi) ={}& \frac{\phi_i\phi_j}{2n} \Gll(p;n) + \left(\delta_{ij}-\frac{\phi_i\phi_j}{2n} \right) \Gtt(p;n) \nonumber \\ & + \eps_{ij} \Glt(p;n) 
\label{Gdef}
\end{align}
for a constant field $\phi$, where 
\beq
\begin{split}
\Gll(p;n) &= - \frac{\Gamma_A(p;n)}{D(p;n)} , \\ 
\Gtt(p;n) &= - \frac{\Gamma_A(p;n)+2n\Gamma_B(p;n)}{D(p;n)} , \\ 
\Glt(p;n) &= \frac{\Gamma_C(p;n)}{D(p;n)},
\end{split}
\label{Gdef_1}
\eeq
and $D=\Gamma_A^2+2n\Gamma_B\Gamma_A+\Gamma_C^2$. Equations (\ref{Gdef_1}) follow from the matrix relation $G^{-1}=-\Gamma^{(2)}$. Using $\lambda,Z_C\sim k^{3-d}\gg k^2$, we then find
\beq
D(p) \simeq 2n_0\lambda V_A(\w^2+c^2\p^2) 
\eeq
and
\beq
\begin{split}
\Gbarll(p) &= - \frac{1}{2n_0\lambda} , \\ 
\Gbartt(p) &= - \frac{1}{V_A} \frac{1}{\w^2+(c\p)^2} = - \frac{2n_0mc^2}{\bar n} \frac{1}{\w^2+(c\p)^2} , \\ 
\Gbarlt(p) &= \frac{Z_C}{2n_0\lambda V_A} \frac{\w}{\w^2+(c\p)^2}  =  \frac{mc^2}{\bar n} \frac{dn_0}{d\mu} \frac{\w}{\w^2+(c\p)^2} .
\end{split}
\label{propa_ir}
\eeq
The propagators $\Gbartt$ and $\Gbarlt$ have well defined limits when $k\to 0$, while the longitudinal propagator $\Gbarll\sim 1/k^{(3-d)}$ diverges in agreement with the general discussion of the Introduction. {\it Stricto sensu}, equations (\ref{propa_ir}) hold in the limit $|\p|,|\w|/c\ll k$. We can nevertheless obtain the propagators at $k=0$ and finite $(\p,i\w)$ by stopping the flow at $k\sim \sqrt{\p^2+\w^2/c^2}$ (see the discussion in Sec.~\ref{subsec_analytic}). Since the local gauge invariant (thermodynamic) quantities are not expected to flow in the infrared limit (Sec.~\ref{sec_nr}), this procedure amounts to replacing $n_0$, $\bar n$, $c$  and $dn_0/d\mu$ by their $k=0$ values. As for the longitudinal correlation function, we reproduce the expected infrared singularity
\beq
\Gbarll(p) = - \frac{1}{2n_0C [\w^2+(c\p)^2]^{(3-d)/2}} .
\label{Gll}
\eeq
The constant $C$ can be estimated by comparing (\ref{Gll}) with the result of Popov's hydrodynamic theory~\cite{Giorgini92},
\beq
C \simeq \left(\frac{2\bar n}{mcn_0}\right)^2 .
\label{C_est}
\eeq
From these results, we deduce the infrared behavior of the normal and anomalous propagators
\beq
\begin{split} 
G_{\rm n}(p) ={}& - \mean{\psi(p)\psi^*(p)} \\
={}& - \frac{n_0mc^2}{\bar n} \frac{1}{\w^2+(c\p)^2} \\ & - \frac{mc^2}{\bar n} \frac{dn_0}{d\mu} \frac{i\w}{\w^2+(c\p)^2} + \half \Gbarll(p) , \\
G_{\rm an}(p) ={}& - \mean{\psi(p)\psi(-p)} \\
={}& \frac{n_0mc^2}{\bar n} \frac{1}{\w^2+(c\p)^2} + \half \Gbarll(p) .
\end{split}
\label{Gnan}
\eeq
The leading order terms in (\ref{Gnan}) agree with the results of Gavoret and Nozi\`eres~\cite{Gavoret64}. The contribution of the diverging longitudinal correlation function was first identified by NN, and later in Refs.~\cite{Weichman88,Giorgini92,Castellani97,Pistolesi04}.

\section{RG equations}
\label{sec_rg}

To compute approximately the effective potential $U$ and the one-particle propagator, we follow the BMW NPRG scheme proposed in Refs.~\cite{Blaizot06,Benitez08,Benitez09} with a truncation in fields to lowest non-trivial order~\cite{Guerra07}. 

\subsection{BMW equations}

The dependence of the effective action on $k$ is given by Wetterich's equation~\cite{Wetterich93}
\beq
\partial_t \Gamma[\phi] = \half \Tr \left\lbrace \dot
R\bigl(\Gamma^{(2)}[\phi]+R\bigr)^{-1}\right\rbrace ,
\label{Weq} 
\eeq
where $t=\ln(k/\Lambda)$ and $\dot R=\dt R$. In Fourier space, the trace involves a sum over frequencies and momenta as well as a trace over the two components of the field $\phi=(\phi_1,\phi_2)$.

The flow equation for the effective potential $U(n)=(\beta V)^{-1}\Gamma[\phi]$ (with $\phi=(\sqrt{2n},0)$ is directly derived from (\ref{Weq}), 
\beq
\dt U(n) = - \half \int_p \dot R(p) [\Gll(p;n)+\Gtt(p;n)] , 
\label{flowU}
\eeq
where
\beq
\int_p = \int_\p \int_\w = \int \frac{d^dp}{(2\pi)^d} \intinf \frac{d\w}{2\pi}. 
\eeq 
The flow equation of the condensate density is then deduced from 
\beq
\dt U'(n_0)=\dt U'|_{n_0}+U''(n_0) \dt n_0 = 0 , 
\eeq
while that of $\lambda=U''(n_0)$ is obtained from 
\beq
\dt \lambda = \dt U''|_{n_0} + U'''(n_0) \dt n_0 .
\eeq
Note that the propagator $G$ in (\ref{flowU}) and below is defined as the inverse of $-(\Gamma^{(2)}+R)$.

Equation (\ref{Weq}) leads to a flow equation for the two-point vertex $\Gamma^{(2)}$ which involves the three-point and four-point vertices,
\begin{multline}
\dt \Gamma^{(2)}_{ij}(p;\phi) = \\ 
-\half \sum_{q,i_1,i_2} \tilde\dt G_{i_1i_2}(q;\phi) \Gamma^{(4)}_{iji_2i_1}(p,-p,q,-q;\phi)  \\ 
- \half \sum_{q,i_1\cdots i_4} \Bigl\lbrace \Gamma^{(3)}_{ii_2i_3}(p,q,-p-q;\phi) \Gamma^{(3)}_{ji_4i_1}(-p,p+q,-q;\phi) \\
\times [\tilde\dt G_{i_1i_2}(q;\phi)]G_{i_3i_4}(p+q;\phi) + (p\leftrightarrow -p, i\leftrightarrow j) \Bigr\rbrace ,
\label{flow1} 
\end{multline}
where the operator $\tilde\dt=(\partial R/\partial t)\partial_R$ acts only on the $t$ dependence of the regulator $R$. The field $\phi$ is assumed to be uniform and time independent. 

The BMW approximation is based on the following two observations~\cite{Blaizot06}: i) Since the function $\tilde\dt G_{ij}(q;\phi)$ is proportional to $\dt R(q)$, the integral over the loop variable $q=(\q,i\w)$ in (\ref{flow1}) is dominated by values of $|\q|$ and $|\w|/c$ smaller than $k$. (Note that this argument requires a regulator $R(q)$ that acts both on momentum and frequency.). ii) As they are regulated in the infrared, the vertices $\Gamma^{(n)}$ are smooth functions of momenta and frequencies~\cite{note7}. These two properties allow one to make an expansion in power of $\q^2/k^2$ and $\w^2/(ck)^2$, independently of the value of the external variable $p=(\p,i\wnu)$. To leading order, one simply sets $q=0$ in the three- and four-point vertices. We can then obtain a close equation for $\Gamma^{(2)}_{ij}(p;\phi)$ by noting that~\cite{Blaizot06}
\beq
\begin{split}
\Gamma^{(3)}_{ijl}(p,-p,0;\phi) &= \frac{1}{\sqrt{\beta V}} \frac{\partial}{\partial\phi_l} \Gamma^{(2)}_{ij}(p;\phi) , \\ 
\Gamma^{(4)}_{ijlm}(p,-p,0,0;\phi) &= \frac{1}{\beta V} \frac{\partial^2}{\partial\phi_l \partial\phi_m} \Gamma^{(2)}_{ij}(p;\phi) .
\end{split}
\label{gam_bmw} 
\eeq
The flow equation for $\Gamma_\alpha(p;n)$ is given in Appendix \ref{app_floweq} [Eqs~(\ref{eq1_bmw}-\ref{eq3_bmw})]. 

\subsection{Truncated flow equations} 
\label{subsec_truncated} 

We simplify the BMW equations by considering two additional approximations. First we define the self-energy $\Delta_\alpha(p;n)$ ($\alpha=A,B,C$) by
\beq
\begin{split}
\Gamma_A(p;n) &= \eps_\p + U'(n) + \Delta_A(p;n) , \\  
\Gamma_B(p;n) &= U''(n) + \Delta_B(p;n) , \\  
\Gamma_C(p;n) &= \w + \Delta_C(p;n) . 
\end{split}
\label{deltadef}
\eeq
It satisfies $\Delta_\alpha(p=0;n)=0$. We then expand $\Delta_\alpha(p;n)$ about $n_0$,
\beq
\Delta_\alpha(p;n) = \Delta_\alpha(p;n_0) + (n-n_0) \Delta^{(1)}_\alpha(p;n_0) + \cdots, 
\eeq
and truncate the expansion to lowest order, \ie we approximate $\Delta_\alpha(p;n)$ by its value $\bar\Delta_\alpha(p)=\Delta_\alpha(p;n_0)$ in the equilibrium state. Similarly we truncate the effective potential $U$ to second-order, i.e.
\beq
U(n) = U(n_0) + \frac{\lambda}{2} (n-n_0)^2 ,
\label{Utrunc}
\eeq
where $\lambda=U''(n_0)$. For $k=\Lambda$, the effective action is given by the microscopic action $S[\phi]$ (Sec.~\ref{sec_ea}), so that $n_0|_{k=\Lambda}=\mu/g$, $\lambda|_{k=\Lambda}=g$ and $\Delta_\alpha|_{k=\Lambda}=0$ (Bogoliubov approximation).

The second approximation is based on a derivative expansion of the vertices and propagators. We have already pointed out that the integral over the internal loop variable $q$ is dominated by small values $|\q|,|\w|/c\lesssim k$. Furthermore, since the external variable $p=(\p,i\wnu)$ acts as an effective low-energy cutoff, the flow of $\Gamma^{(2)}_{ij}(p;\phi)$ stops when $k$ becomes of the order of $|\p|$ or $|\w|/c$. Thus all propagators and vertices in (\ref{flow1}) should be evaluated in the momentum and frequency range $|\q|,|\p+\q|\lesssim k$ and $|\w|/c,|\w+\wnu|/c\lesssim k$. In addition to the BMW approximation, we can therefore use the derivative expansion (\ref{expansion}) of the vertices in the rhs of (\ref{flow1}). This approximation has been shown be very reliable in classical models~\cite{Benitez08,Sinner08,Dupuis08}. While we also expect a high degree of accuracy in the low-energy limit $p\to 0$, the approximation is more questionable in the high-frequency limit. The high-frequency behavior of the two-point vertex $\Gamma^{(2)}_{ij}(p)$ (and in turn of the propagator $G_{ij}(p)$) follows from the high-frequency behavior of the propagator $G(p+q)$ appearing in (\ref{flow1}). Clearly the derivative expansion does not reproduce the expected high-frequency limit of the propagator. We shall see nevertheless that the solution of the flow equations does not contradict the $\wnu\to\infty$ limit of the propagator (Appendix \ref{app_high_w}) although the leading corrections $\calO(1/\wnu)$ and $\calO(1/\wnu^2)$ are likely to be incorrect.

These two approximations lead to the flow equations (see Appendix \ref{app_floweq}) 
\beq
\begin{split}
\dt n_0 &= \frac{3}{2} \Ibarll + \half \Ibartt , \\
\dt \lambda &= - \lambda^2 \bigl[9\Jbarllll(0) - 6\Jbarltlt(0) + \Jbartttt(0) \bigr] ,
\end{split}
\label{flow5} 
\eeq
and 
\begin{align}
\dt \bar\Delta_A(p) ={}& \lambda \dt n_0 - \frac{\lambda}{2}(\Ibarll+3\Ibartt) \nonumber \\ & -2n_0\lambda^2  \bigl[\Jbarlltt(p)+\Jbarttll(p)+2\Jbarltlt(p)\bigr] , \nonumber \\
\dt \bar\Delta_B(p) ={}& - \dt\lambda + \frac{\lambda}{2n_0}(\Ibartt-\Ibarll) \nonumber \\
& +\lambda^2 \bigl[-9\Jbarllll(p)+\Jbarlltt(p) \nonumber \\ & +\Jbarttll(p)-\Jbartttt(p)+8\Jbarltlt(p) \bigr] , \label{flow3} \\ 
\dt \bar\Delta_C(p) ={}& 2 n_0\lambda^2 \bigl[\Jbarttlt(p) - \Jbarlttt(p)\nonumber  \\ & -3\Jbarlllt(p) + 3\Jbarltll(p) \bigr] . \nonumber 
\end{align}
where the coefficients $\bar J_{\alpha\beta}(p)=J_{\alpha\beta}(p;n_0)$ and $\bar I_\alpha=I_\alpha(n_0)$ are defined by 
\beq
\begin{split}
I_\alpha(n) &= \int_q \tilde\dt G_\alpha(q;n) , \\ 
J_{\alpha\beta}(p;n) &= \int_q [\tilde\dt G_\alpha(q;n)] G_\beta(p+q;n) ,
\end{split}
\label{IJdef}
\eeq
with $\alpha,\beta={\rm ll,tt,lt}$. The flow equations for $Z_A$, $V_A$, and $Z_C$ are simply derived from 
\beq
\begin{split}
\dt Z_A &= \frac{\partial}{\partial\eps_\p} \dt\Gamma_A(p;n)\Bigl|_{n=n_0,p=0} , \\ 
\dt V_A &= \frac{\partial}{\partial\w^2} \dt\Gamma_A(p;n)\Bigl|_{n=n_0,p=0} , \\ 
\dt Z_C &= \frac{\partial}{\partial\w} \dt\Gamma_C(p;n)\Bigl|_{n=n_0,p=0} .
\end{split}
\label{flow4} 
\eeq
This gives
\beq
\begin{split}
\dt Z_A ={}& -2n_0\lambda^2 \frac{\partial}{\partial\eps_\p} \bigl[ \Jbarlltt(p)+\Jbarttll(p)+2\Jbarltlt(p)\bigr]_{p=0} ,  \\ 
\dt V_A ={}& -2n_0\lambda^2 \frac{\partial}{\partial\w^2} \bigl[ \Jbarlltt(p)+\Jbarttll(p)+2\Jbarltlt(p)\bigr]_{p=0} ,  \\ 
\dt Z_C ={}& 2 n_0\lambda^2 \frac{\partial}{\partial\w} \bigl[\Jbarttlt(p) - \Jbarlttt(p) \\ & -3\Jbarlllt(p) + 3\Jbarltll(p) \bigr]_{p=0} .
\end{split}
\label{flow6} 
\eeq
Equations (\ref{flow5}) and (\ref{flow6}) agree with those obtained from a simple truncation of the effective action $\Gamma[\phi]$~\cite{Dupuis07}. 

\subsection{Analytical solution in the infrared limit} 
\label{subsec_analytic} 

It is convenient to write the flow equations in terms of dimensionless variables 
\beq
\begin{split} 
\tilde n_0 &= k^{-d} Z_C n_0 , \\
\tilde\lambda &= k^d\eps_k^{-1} (Z_AZ_C)^{-1} \lambda \\ 
\tilde V_A &= \eps_k Z_A Z_C^{-2} V_A 
\end{split} 
\label{dimless_def}
\eeq
(see Appendix \ref{app_dimensionless}). In the infrared limit $k\to 0$, the RG equations simplify,
\beq
\begin{split}
\dt\tilde n_0 &= -(d+\eta_C)\tilde n_0 , \\ 
\dt\tilde\lambda &= (d-2+\eta_C)\tilde\lambda + 8 \frac{v_{d+1}}{d+1} \frac{\tilde\lambda^2}{\tilde V_A^{1/2}} , \\ 
\eta_C &= - 8 \frac{v_{d+1}}{d+1} \frac{\tilde\lambda}{\tilde V_A^{1/2}} , \\ 
\dt\tilde V_A &= (2+2\eta_C) \tilde V_A, 
\end{split}
\label{flow7} 
\eeq
where $\eta_C=-\dt\ln Z_C$ (see Appendix \ref{app_flow_ir}). We deduce
\beq
\dt\tilde\lambda = (d-2)\tilde\lambda
\eeq
and
\beq
\dt \eta_C = (d-3)\eta_C - \eta_C^2 . 
\eeq
For $d=3$, one finds
\beq
\eta_C = \frac{\eta_C^0}{1+\eta_C^0 t} ,
\label{etaC1}
\eeq
with $\eta_C^0$ a constant, whereas
\beq
\eta_C \to d-3
\label{etaC2}
\eeq
for $d<3$. The asymptotic behavior deduced from (\ref{etaC1}) and (\ref{etaC2}) is summarized in Table \ref{table}. In particular one finds that the coupling constant $\lambda$ vanishes when $k\to 0$, $\lambda\sim (\ln k)^{-1}$ for $d=3$ and $\lambda\sim k^{3-d}$ for $d<3$, in agreement with the expected divergence of the longitudinal correlation function (Sec.~\ref{sec_de}). 

\begin{table}
\renewcommand{\arraystretch}{1.5}
\begin{center}
\begin{tabular}{|c||c|c|}
\hline 
& $d=3$ & $1<d<3$ 
\\ \hline \hline 
$n_0$ & $n_0^*$ & $n_0^*$ 
\\ \hline 
$\lambda$ & $(\ln k)^{-1}$ & $k^\eps$ 
\\ \hline 
$Z_C$ & $(\ln k)^{-1}$ & $k^\eps$
\\ \hline
$V_A$ & $V_A^*$ & $V_A^*$ 
\\ \hline \hline
$\tilde n_0$ & $(k^3\ln k)^{-1}$ & $k^{2\eps-3}$ 
\\ \hline
$\tilde\lambda$ & $k$ & $k^{1-\eps}$ 
\\ \hline
$\tilde V_A$ & $(k\ln k)^2$ & $k^{2-2\eps}$ 
\\ \hline \hline
$\tilde n_0'$ & $k^{-2}$ & $k^{\eps-2}$ 
\\ \hline 
$\tilde \lambda'$ & $(\ln k)^{-1}$ & $\tilde \lambda'{}^*$ 
\\ \hline 
\end{tabular}
\end{center}
\caption{Asymptotic behavior for $k\to 0$ ($\eps=3-d$). The stared quantities
  indicate nonzero fixed-point values. } 
\label{table}
\end{table}

In the infrared limit, $Z_C$ is suppressed (Table \ref{table}) and does not play any role in the leading behavior for $k\to 0$ [Eqs.~(\ref{flow7})]. Discarding $Z_C$, we two-point vertex $\Gamma^{(2)}$ exhibits a space-time $SO(d+1)$ (relativistic) symmetry. It is possible to eliminate the anisotropy between time and space by rescaling the frequency, $\tilde\w\to \tilde\w\tilde V_A^{-1/2}$ (the dimensionless frequency $\tilde\w$ is defined in Appendix \ref{IJ_dim}). To maintain the dimensionless form of the effective action, one should also rescale the (dimensionless) field, $\tilde\phi\to \tilde V_A^{-1/4}\tilde\phi$. This leads to an isotropic relativistic model with dimensionless condensate density and coupling constant defined by
\beq
\tilde n_0' = \sqrt{\tilde V_A} \tilde n_0 , \qquad 
\tilde\lambda' = \frac{\tilde\lambda}{\sqrt{\tilde V_A}} .
\eeq
The asymptotic behavior of $\tilde n_0'$ and $\tilde\lambda'$ is in agreement with the known results of the classical $(d+1)$-dimensional $O(2)$ model (table \ref{table}). In particular, the dimensionless coupling constant $\tilde\lambda'$ vanishes logarithmically when $d+1=4$ and reaches a non-zero fixed point value $\tilde\lambda'{}^*$ when $d+1<4$. Using
\beq
\lambda = k^{-d} (Z_A\eps_k)^{3/2} V_A^{1/2} \tilde\lambda' \sim k^{3-d} \tilde\lambda',
\eeq
we deduce that $\lambda$ vanishes as $k^{3-d}$ when $d<3$ and logarithmically when $d=3$. Thus, the divergence of the longitudinal susceptibility (which follows from the vanishing of $\lambda$) can be understood as a consequence of the low-energy behavior of the classical $(d+1)$-dimensional $O(2)$ model. 

As explained in Appendix \ref{app_flow_ir}, the infrared limit of the self-energies can be obtained from the derivative expansion if we stop the flow at $k\sim (\p^2+\w^2/c^2)^{1/2}$. This yields
\beq
\begin{split} 
\bar\Gamma_{A,k=0}(p) &\simeq V_A \w^2 + Z_A \eps_\p , \\ 
\bar\Gamma_{B,k=0}(p) &\sim (\w^2+c^2\p^2)^{(3-d)/2} , \\ 
\bar\Gamma_{C,k=0}(p) &\sim  \w (\w^2+c^2\p^2)^{(3-d)/2} .
\end{split}
\label{gam_ir1} 
\eeq
Since $Z_A$ and $V_A$ do not flow when $k\to 0$, they can be evaluated for $k=0$ and related to thermodynamic quantities (Sec.~\ref{sec_de}). We expect the following relation between $\bar\Gamma_B$ and $\bar\Gamma_C$,
\beq
\lim_{p\to 0} \frac{\bar\Gamma_{C,k=0}(p)}{\w \bar\Gamma_{B,k=0}(p)} = \lim_{k\to 0} \frac{Z_C}{\lambda} = \frac{dn_0}{d\mu}\biggl|_{k=0} .
\label{gam_ir2} 
\eeq
This relation will be confirmed numerically in Sec.~\ref{sec_nr}. From (\ref{gam_ir1}) and (\ref{gam_ir2}), we reproduce the infrared limit (\ref{propa_ir},\ref{Gll}) of the propagators obtained in Sec.~\ref{subsec_1PP} from general considerations.

\section{Numerical results} 
\label{sec_nr}

In this section we discuss the numerical solution of the flow equations. We consider a two-dimensional system in the weak coupling limit $2mg=0.1$. The actual boson density $\bar n\equiv \bar n_{k=0}$ is fixed and the chemical potential $\mu=gn_{0,k=\Lambda}$ is fine tuned in order to obtain $n_{s,k}=Z_{A,k}n_{0,k}=\bar n$ for $k=0$. 

\begin{figure}
\centerline{\includegraphics[bb=173 225 509 463,width=7cm,clip]{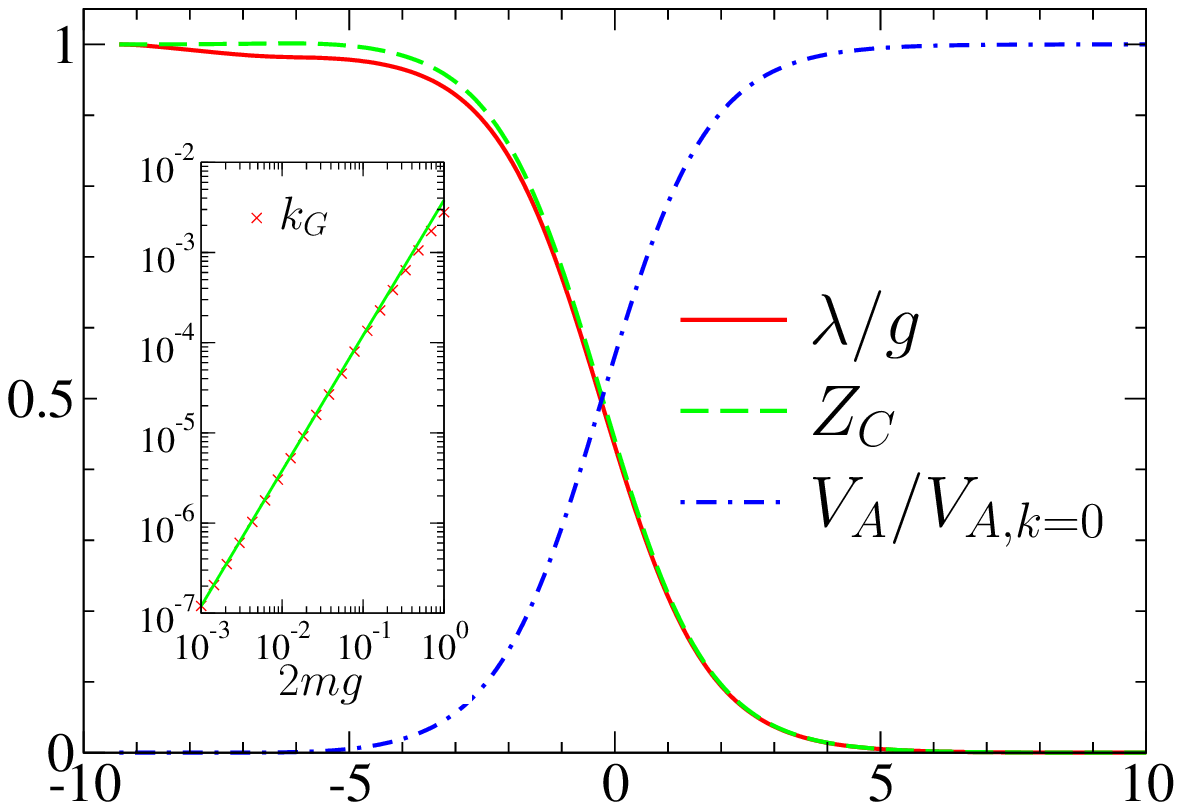}}
\centerline{\includegraphics[width=7cm,clip]{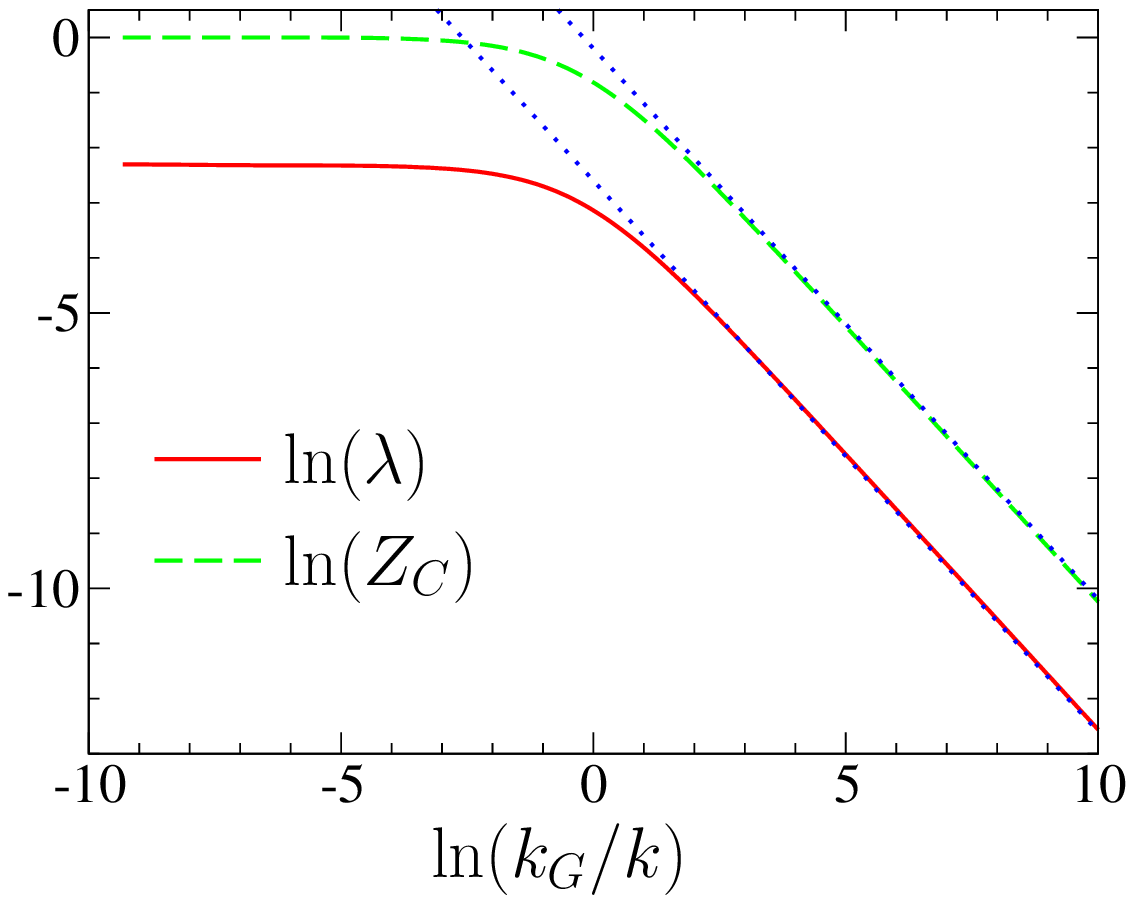}}
\caption{(Color online) Top panel: $\lambda/g$, $Z_C$ and $V_A/V_{A,k=0}$ vs $\ln(k_G/k)$ where $k_G=\sqrt{(gm)^3\bar n}/4\pi$, for $\bar n=0.01$, $2mg=0.1$ and $d=2$ [$\ln(k_G/k_h)\simeq -5.87$]. The inset shows $k_G$ vs $2mg$ obtained from the criterion $V_{A,k_G}=V_{A,k=0}/2$ [the green solid line is a fit to $k_G \propto (2mg)^{3/2}$]. Bottom panel: $\ln(\lambda)$ and $\ln(Z_C)$ vs $\ln(k_G/k)$. $\lambda$ and $Z_C$ vanish as $k$ for $k\ll k_G$ (blue dotted lines). All figures are obtained with $\Lambda=2m=1$ and the regulator (\ref{regdef}).}
\label{fig_lambda}
\end{figure}

\begin{figure}
\centerline{\includegraphics[width=6cm,clip]{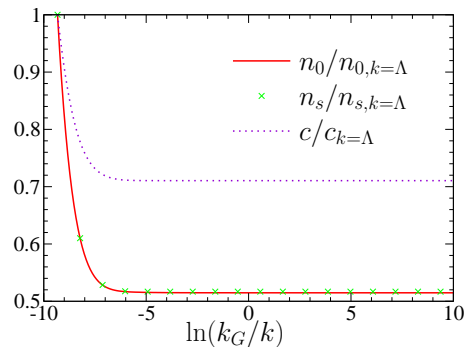}}
\caption{(Color online) Condensate density $n_0$, superfluid density $n_s=Z_An_0=\bar n$ and Goldstone mode velocity $c$ vs $\ln(k_G/k)$. The parameters are the same as in Fig.~\ref{fig_lambda}.}
\label{fig_density}
\end{figure}

\begin{figure}
\centerline{\includegraphics[width=6cm,clip]{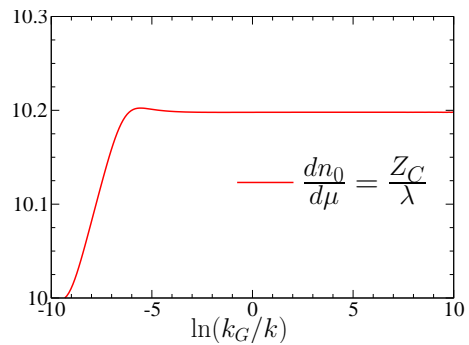}}
\caption{(Color online) Condensate compressibility $dn_0/d\mu=Z_C/\lambda$ vs $\ln(k_G/k)$ [Eq.~(\ref{ward1})].} 
\label{fig_kappa0} 
\end{figure}

The flow of $\lambda$, $Z_C$ and $V_A$ is shown in Fig.~\ref{fig_lambda}. (The asymptotic behavior of various quantities as a function of the space dimension is summarized in table \ref{table}.) In agreement with the discussion of Secs.~\ref{sec_de} and \ref{sec_rg}, we find that $\lambda,Z_C\sim k$ are suppressed as $k\to 0$, while $V_A$ flows toward a finite value. The anomalous self-energy $\Sigan(p=0)=n_0\bar\Gamma_B(0)=n_0\lambda$ therefore vanishes for $k\to 0$ in agreement with the exact result~\cite{Nepomnyashchii75}. From Fig.~\ref{fig_lambda}, one can clearly identify the momentum scale $k_G$ below which the Bogoliubov approximation breaks down. The inset in the figure shows $k_G$ obtained from the criterion $V_{A,k_G}=V_{A,k=0}/2$. It is proportional to $\sqrt{(gm)^3\bar n} \sim gm k_h\ll k_h$, in agreement with the perturbative estimate (\ref{kg_est}). In practice, we use the definition $k_G=\sqrt{(gm)^3\bar n}/4\pi$. Note that the healing scale $k_h=\sqrt{2mg\bar n}$ (defined in Appendix \ref{app_bog}) keeps its mean-field (Bogoliubov) expression since the renormalization of the two-point vertex is very small for $k\sim k_h\gg k_G$. 

Fig.~\ref{fig_density} shows the behavior of the thermodynamic quantities $n_0$, $n_s$ and $c$. Since $Z_{A,k=0}\simeq 1.004$, the mean boson density $\bar n=Z_An_0$ is nearly equal to the condensate density $n_0$. The condensate compressibility $dn_0/d\mu=Z_C/\lambda$ [Eq.~(\ref{ward1})] is shown in Fig.~\ref{fig_kappa0}. Apart from an initial variation at the beginning of the flow ($k\lesssim k_h$), these quantities do not vary with $k$. In particular, they are not sensitive to the Ginzburg scale $k_G$. This result is particularly remarkable for the Goldstone mode velocity $c$, whose expression (\ref{cdef}) involves the parameters $\lambda$, $Z_C$ and $V_A$, which all strongly vary when $k\sim k_G$. These findings are a nice illustration of the fact that the divergence of the longitudinal susceptibility does not affect local gauge invariant quantities~\cite{Pistolesi04}. 

\begin{figure}
\centerline{\includegraphics[width=7cm,clip]{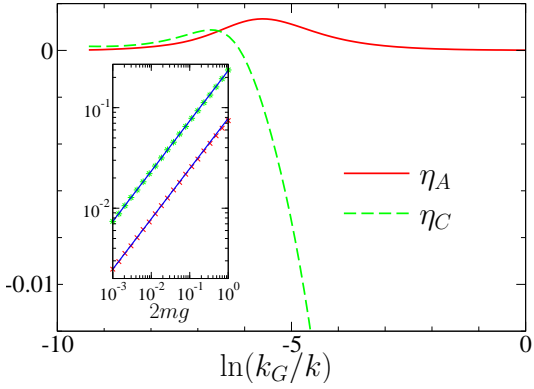}}
\caption{(Color online) $\eta_A$ and $\eta_C$ vs $\ln(k_G/k)$. The inset shows the location of the maxima in the curves $\eta_A$ (red crosses) and $\eta_C$ (green stars) vs $2mg$ [the blue solid lines correspond to $\const\times \sqrt{2mg\bar n}\propto k_h$]. The parameters are the same as in Fig.~\ref{fig_density}.}
\label{fig_eta}
\end{figure}

In Fig.~\ref{fig_eta} we show the flow of $\eta_A=-\dt \ln Z_A$ and $\eta_C=-\dt \ln Z_C$ for $k>k_G$. Both $\eta_A$ and $\eta_C$ exhibit a maximum corresponding to a slight increase of $Z_A$ and $Z_C$ ($Z_A$ then saturates to $Z_{A,k=0}$ while $Z_C$ strongly decreases when $k\sim k_G$). The location of these maxima is given by the healing scale $k_h$ (see inset in Fig.~\ref{fig_eta})~\cite{note3}. The maxima of $\eta_A$ and $\eta_C$ become more pronounced as $2mg$ increases, but remains very small in the weak coupling limit $2mg\lesssim 1$. The small window around $k_h$ where the anomalous dimension $\eta_A$ is finite is likely to be a remnant of the critical regime that exists near the Goldstone regime at higher temperatures, and which is progressively suppressed as the temperature decreases.

\subsection{Self-energies}

The self-energies are obtained from the numerical solution of the flow equations (\ref{flow3}) or (\ref{flow_dim}). By computing $\bar\Delta_\alpha(\p,i\w)$ for $N$ frequency points $i\w_l$ ($l=1,\cdots,N$ with typically $N\sim 100$), one can construct a $N$-point Pad\'e approximant $P_\alpha(\p,z)$ which is equal to $\bar\Delta_\alpha(\p,i\w)$ when the complex frequency $z$ coincides with one of the Matsubara frequencies $i\w_l$. The retarded part of the self-energy is then approximated by $\bar\Delta^R_\alpha(\p,\w)=P_\alpha(\p,\w+i0^+)$~\cite{Vidberg77}. (All self-energies discussed in this section corresponds to $k=0$.)

\begin{figure}
\centerline{\includegraphics[width=7cm,clip]{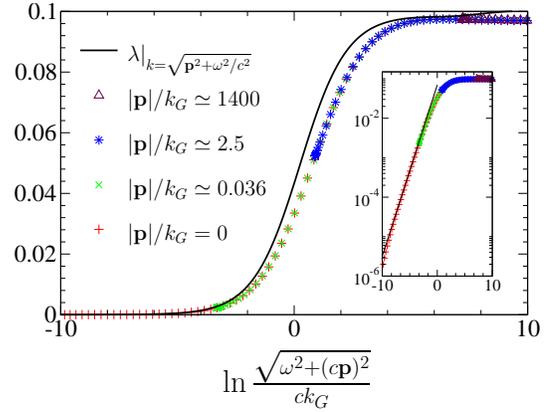}}
\caption{(Color online) $\bar\Gamma_{B,k=0}(\p,i\w)=\bar\Delta_{B,k=0}(\p,i\w)$ (symbols) and $\lambda_{k=\sqrt{\p^2+\w^2/c^2}}$ (solid line) vs $\sqrt{\w^2+(c\p)^2}$ for various values of $|\p|$. The inset shows $\ln(\bar\Gamma_{B,k=0})$ and a fit to $C\sqrt{\w^2+(c\p)^2}$ (solid line).}
\label{fig_gamB_2}
\end{figure}

\begin{figure}
\centerline{\includegraphics[width=6cm,clip]{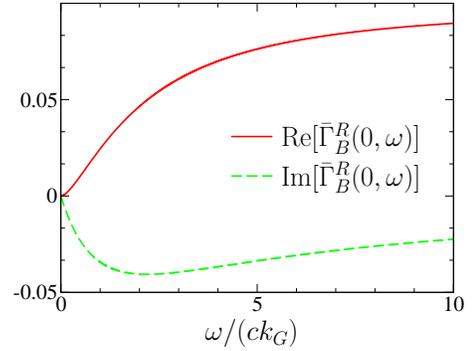}}
\caption{(Color online) Real and imaginary parts of the retarded vertex $\bar\Gamma^R_{B,k=0}(\p=0,\w)$ vs $\w/(ck_G)$.}
\label{fig_gamB_5} 
\end{figure}

Let us first discuss the momentum and frequency dependence of $\bar\Delta_B(p)$ at $k=0$. Note that $\bar\Delta_{B}(p)=\bar\Gamma_{B}(p)=\Sigan(p)/n_0$ since $\lambda_{k=0}=0$. In the following, we shall rather discuss $\bar\Gamma_B(p)$ which is the right quantity to consider when comparing to the Bogoliubov approximation. Fig.~\ref{fig_gamB_2} shows that $\bar\Gamma_B(p)$ is a function of $\w^2+(c\p)^2$, not only in the infrared regime $\sqrt{\p^2+\w^2/c^2}\ll k_G$ but also for $\sqrt{\p^2+\w^2/c^2}>k_G$. Furthermore, $\bar\Gamma_B(p)$ is related to the running coupling constant $\lambda_k$ by 
\beq
\bar\Gamma_B(p) \simeq \lambda|_{k=\sqrt{\p^2+\w^2/c^2}} .
\eeq
This confirms that $\bar\Gamma_B(p)$ can be (approximately) obtained from $\bar\Gamma_{B,k}(p=0)$ by stopping the flow at $k\sim \sqrt{\p^2+\w^2/c^2}$. For $\sqrt{\p^2+\w^2/c^2}\gg k_G$, one therefore recovers the Bogoliubov result $\bar\Gamma_B(p)\simeq g$, while for $\sqrt{\p^2+\w^2/c^2}\ll k_G$ one obtains 
\beq
\bar\Gamma_B(p)\simeq C\sqrt{\w^2+(c\p)^2} ,
\label{nr1}
\eeq
with $C$ a $\p$-independent constant. 

\begin{figure}
\centerline{\includegraphics[width=7cm,clip]{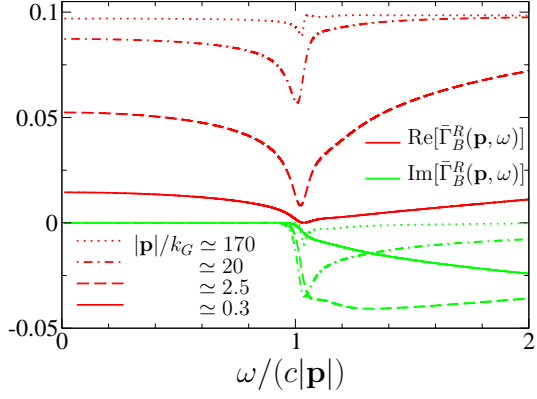}}
\caption{(Color online) Real and imaginary parts of the retarded vertex $\bar\Gamma^R_{B,k=0}(\p,\w)=\Sigma^R_{\rm an}(\p,\w)/n_0$ for various values of $|\p|$ ranging from $0.3 k_G$ up to $170k_G\sim k_h/2$ ($\bar n=0.01$ and $2mg=0.1$). The Bogoliubov approximation corresponds to $\bar\Gamma_B^R(\p,\w)=g=0.1$. ($\Re[\bar\Gamma_B^R]\geq 0$ and  $\Im[\bar\Gamma_B^R]\leq 0$.)}
\label{fig_gamB_3}
\end{figure}

\begin{figure}
\centerline{\includegraphics[width=6cm,clip]{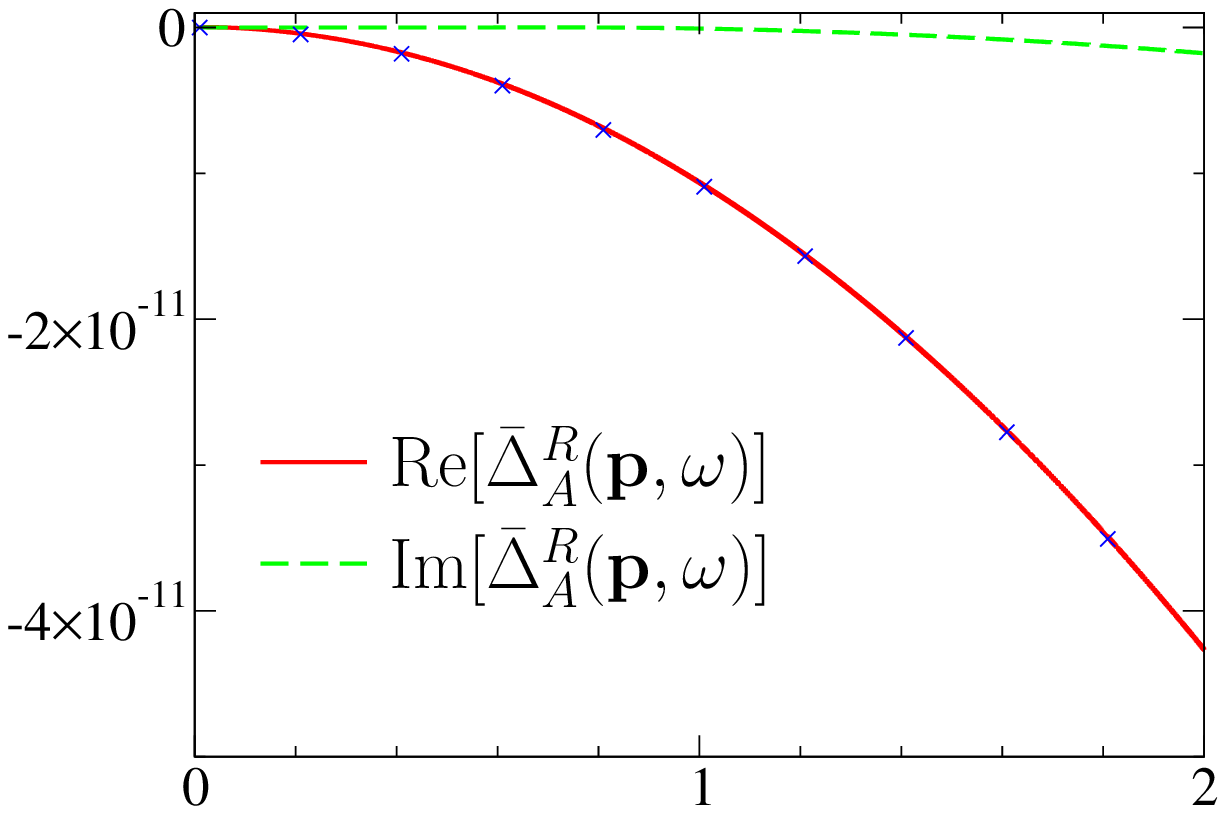}}
\centerline{\includegraphics[width=6cm,clip]{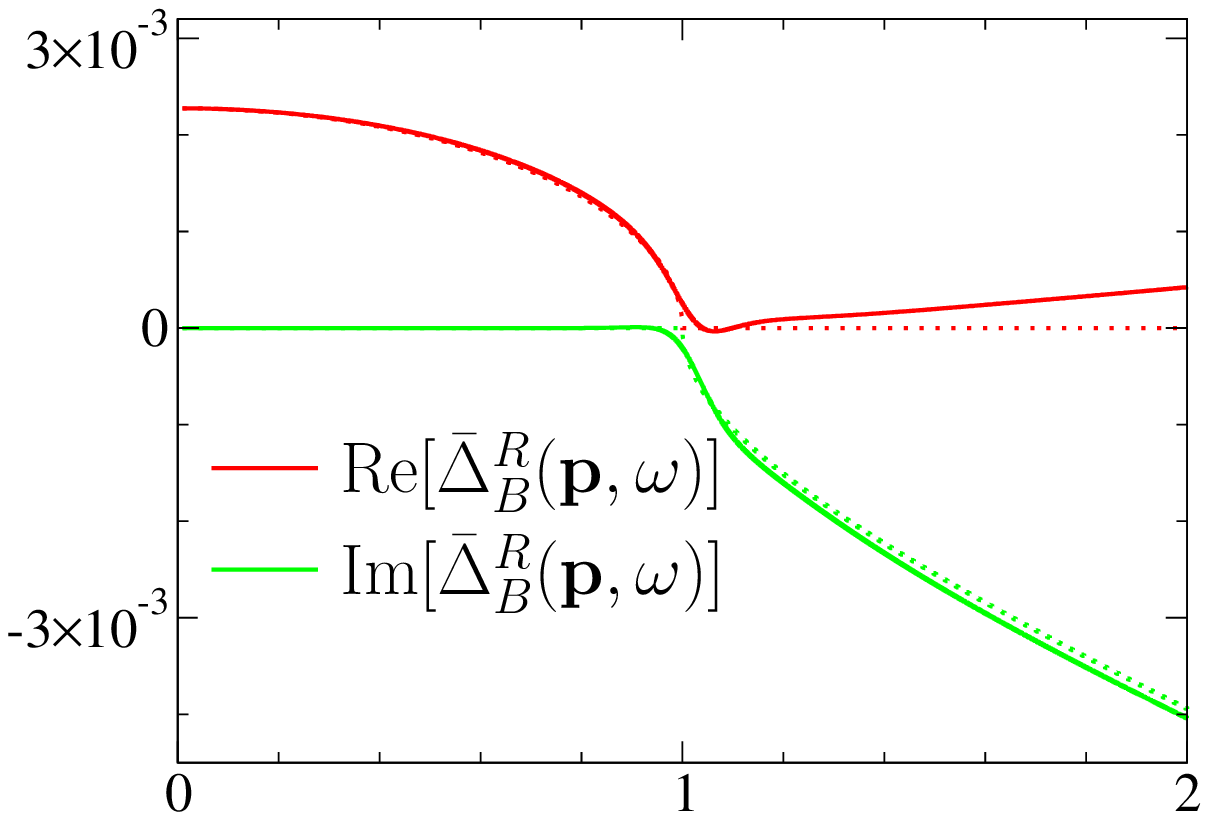}}
\centerline{\includegraphics[width=6cm,clip]{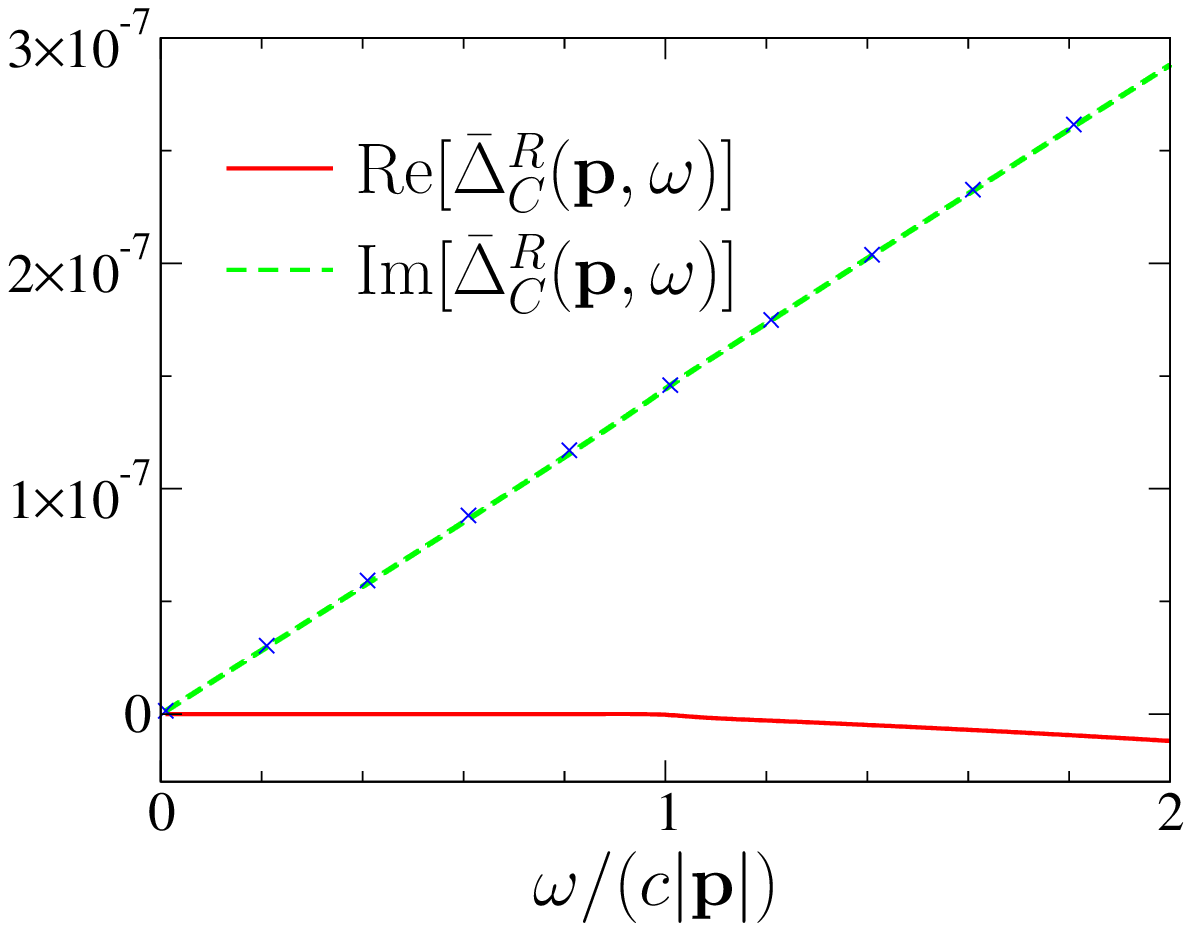}}
\caption{(Color online) Real and imaginary parts of the retarded self-energies $\bar
\Delta^R_A(\p,\w)$, $\bar\Delta^R_B(\p,\w)=\bar\Gamma^R_B(\p,\w)$ and $\bar\Delta^R_C(\p,\w)$ for $|\p|\simeq 0.036 k_G$ and $k=0$. The blue crosses correspond to the expressions (\ref{nr3}) obtained from the derivative expansion, while the dotted lines show the analytical result (\ref{nr2}) obtained from the approximation (\ref{nr1}).}
\label{fig_gamABC}
\end{figure}

\begin{figure}
\centerline{\includegraphics[width=6cm,clip]{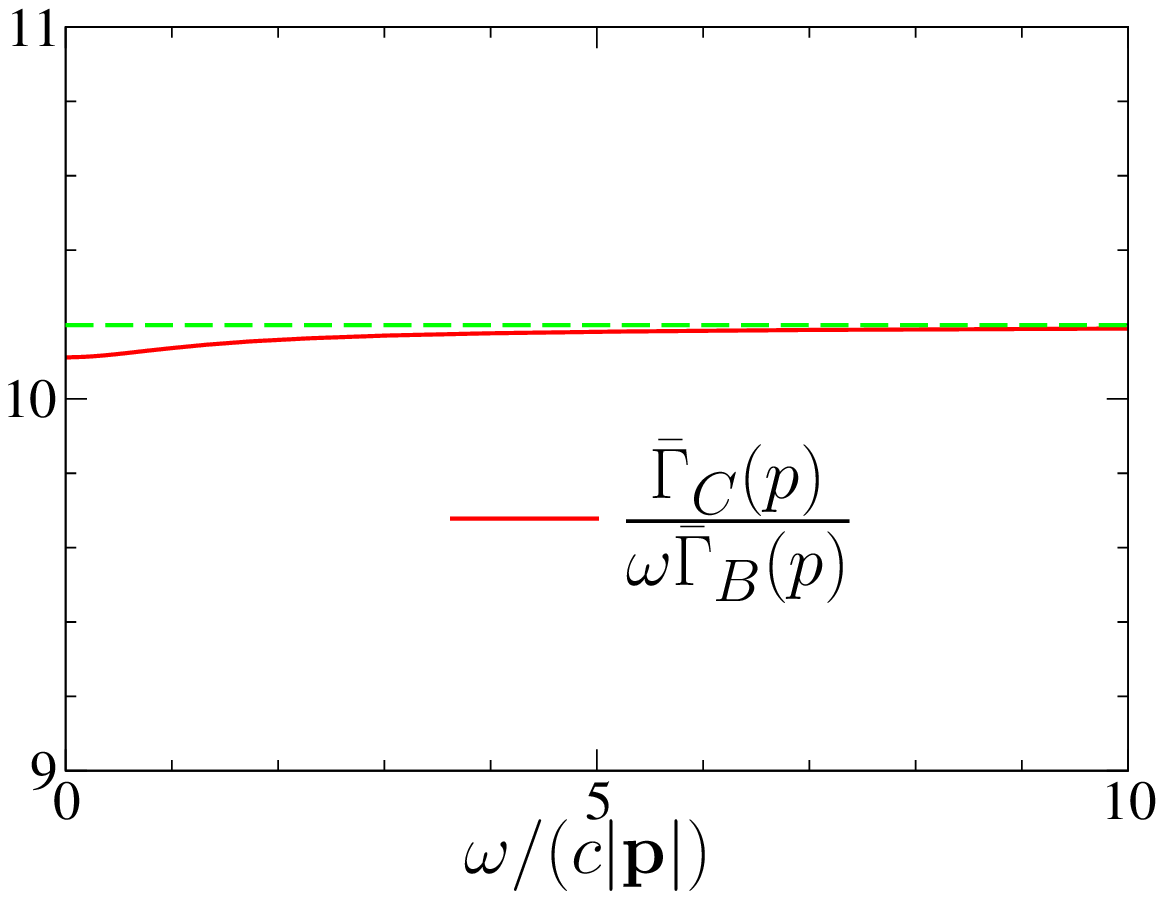}}
\caption{(Color online) $\bar\Gamma_C(p)/(\w\bar\Gamma_B(p))$ vs $\w/(c|\p|)$ for $|\p|\simeq 0.036 k_G$ and $k=0$. The dashed line shows the value of $dn_0/d\mu|_{k=0}=\lim_{k\to 0}Z_C/\lambda$.}
\label{fig_gamC_ratio}
\end{figure}

The Ginzburg scale $k_G$ manifests itself also in the frequency dependence of the retarded vertex $\bar\Gamma^R_B(\p=0,\w)$ (Fig.~\ref{fig_gamB_5}). For $|\w|\gg ck_G$, the imaginary part $\Im[\bar\Gamma^R_B(0,\w)]$ is very small and the real part tends to $g$ in agreement with the Bogoliubov approximation and the exact high-frequency limit (Appendix \ref{app_high_w}). But for $|\w|\ll ck_G$, the real part is strongly suppressed and becomes of the same order as the imaginary part. 
The crossover between the Bogoliubov and the infrared regimes can also be observed by varying $|\p|$ (Fig.~\ref{fig_gamB_3}). While the Bogoliubov result $\bar\Gamma^R_B(\p,\w)=g$ is a good approximation when $|\p|\gg k_G$, $\bar\Gamma^R_B(\p,\w)$ develops a strong frequency dependence for $|\p|\lesssim k_G$. For $|\p|\ll k_G$, we can use (\ref{nr1}) to obtain the low-frequency behavior ($|\w|\ll ck_G$) 
\begin{align}
\bar\Gamma^R_B(\p,\w) \simeq{}& C \sqrt{-(\w+i0^+)^2+(c\p)^2} \nonumber \\ 
\simeq {}& C \theta(c|\p|-|\w|) \sqrt{(c\p)^2-\w^2} 
\nonumber \\ & - i C \sgn(\w) \theta(|\w|-c|\p|) \sqrt{\w^2-(c\p)^2} 
\label{nr2} 
\end{align}
($\theta(x)$ is the step function). The Bogoliubov result $\bar\Gamma_B^R(\p,\w)=g$ is nevertheless reproduced for $|\w|\gg ck_G$ (Fig.~\ref{fig_gamB_5}). As shown in Fig.~\ref{fig_gamABC}, the square-root singularity (\ref{nr2}) is also obtained from the numerical result based on the Pad\'e approximant. The asymptotic result (\ref{nr2}) was first obtained by NN within a diagrammatic approach, and later reproduced by Popov and Seredniakov in the hydrodynamic approach~\cite{Popov79}. Fig.~\ref{fig_gamABC} also shows the numerical results for $\bar\Delta_A^R(\p,\w)$ and $\bar\Delta^R_C(\p,\w)$. In the infrared regime $|\p|,|\w|/c\ll k_G$, these self-energies are very well approximated by their derivative expansion,
\beq
\begin{split}
\bar\Delta^R_A(\p,\w) &\simeq -V_A\w^2 + (Z_A-1)\eps_\p , \\ 
\bar\Delta^R_C(\p,\w) &\simeq i\w , 
\end{split}
\label{nr3}
\eeq
where $V_A\equiv V_{A,k=0}$ and $Z_A\equiv Z_{A,k=0}$. The leading correction to $\bar\Delta^R_C(\p,\w)\simeq i\w$ is given by the relation (\ref{gam_ir2}) between $\bar\Gamma_B(p)$ and $\bar\Gamma_C(p)$, which is rather well satisfied  when $|\p|,|\w|/c\ll k_G$ (Fig.~\ref{fig_gamC_ratio}) 

\begin{figure}
\centerline{\includegraphics[width=6cm,clip]{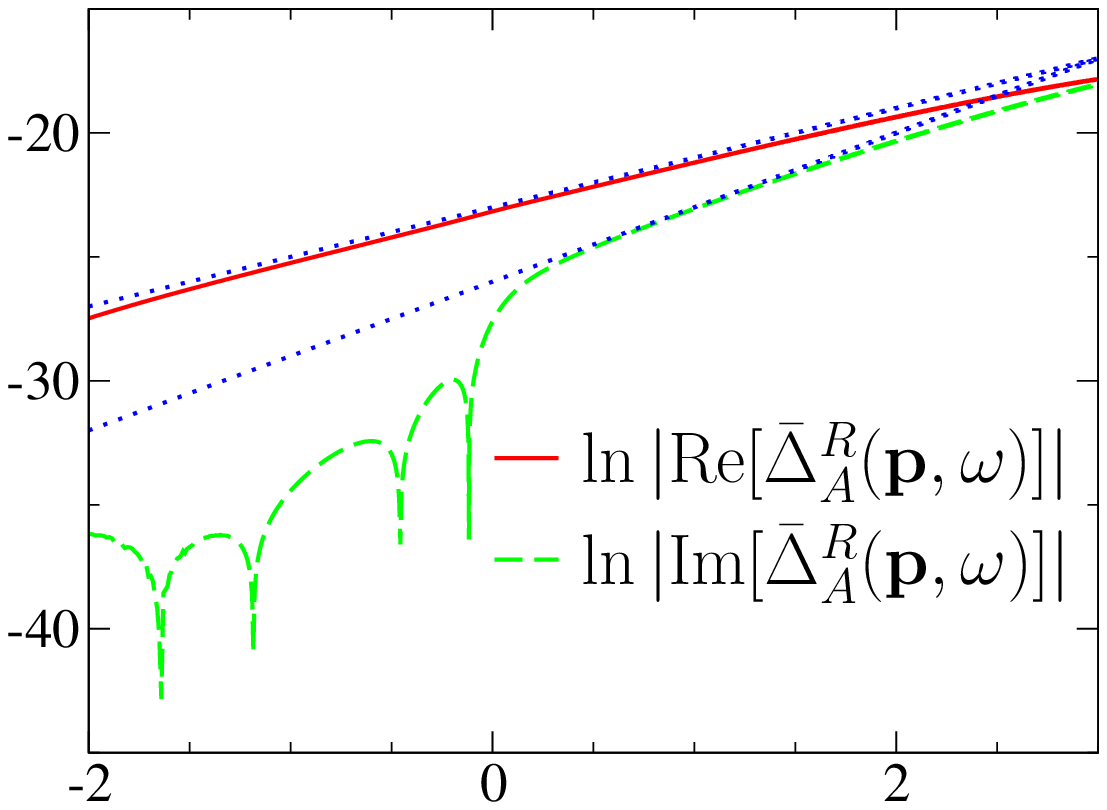}}
\centerline{\includegraphics[width=6cm,clip]{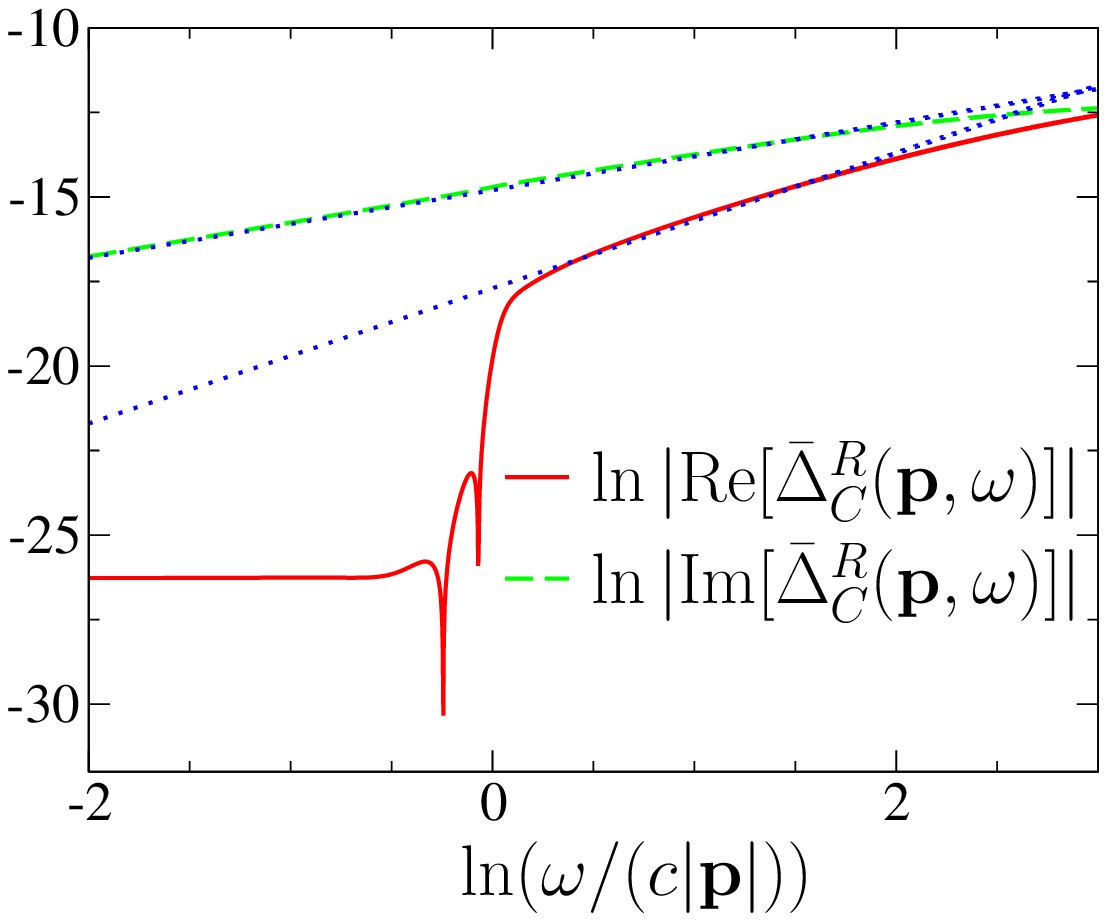}}
\caption{(Color online) Real and imaginary parts of the retarded self-energies $\bar\Delta_A^R(\p,\w)$ and $\bar\Delta_A^R(\p,\w)$. For $\w\gtrsim c|\p|$, $\Re[\bar\Delta_A^R(\p,\w)]\sim \w^2$, $\Im[\bar\Delta_A^R(\p,\w)]\sim \w^3$, $\Re[\bar\Delta_C^R(\p,\w)]\sim \w^2$, $\Im[\bar\Delta_C^R(\p,\w)]\sim \w$ (blue dotted lines). The spikes are due to $\Im[\bar\Delta_A^R(\p,\w)]$ and $\Re[\bar\Delta_C^R(\p,\w)]$ being nearly zero and changing sign.}
\label{fig_damping}
\end{figure}

The imaginary part of $\bar\Delta_A^R$ and the real part of $\bar\Delta_C^R$ give a finite life-time to the sound mode. They arise from the decay of a phonon with momentum $\p$ into two phonons with momenta $\q$ and $\p-\q$ (Beliaev damping~\cite{Beliaev58b}). This damping process follows from the second contribution (proportional to $\bar\Gamma^{(3)}\bar\Gamma^{(3)}$) to $\dt\bar\Gamma^{(2)}$ [Eq.~(\ref{flow1})]. Fig.~\ref{fig_damping} shows that $\Im[\bar\Delta_A^R(\p,\w)]$ and $\Re[\bar\Delta_C^R(\p,\w)]$ vanish for $|\w|\lesssim c|\p|$. The absence of damping below the threshold frequency $\sim c|\p|$ is due to the energy conservation $\w=c|\q|+c|\p-\q|$ in the decay process. While it appears difficult to decide from the numerical results whether $\Im[\bar\Delta_A^R(\p,c|\p|)]$ and $\Re[\bar\Delta_C^R(\p,c|\p|)]$ (which determine the life-time of a phonon with momentum $\p$ and energy $c|\p|$) are zero or not, it is well known that for quasi-particles with a linear spectrum, Beliaev damping cannot take place as there is no phase space available~\cite{Lifshitz_stat_phys_II}. Beliaev damping requires a positive curvature of the quasi-particle dispersion, \ie $E_\p=c|\p|+a|\p|^3$ ($a>0$). In this case, the threshold frequency, obtained from the condition $\w=E_\q+E_{\p-\q}$ (with $\p$ fixed), lies below $E_\p$. The decay of a quasi-particle into a pair of quasi-particles then gives a scattering rate of order $|\p|^3$ in a two-dimensional system~\cite{Kreisel08,Chung08}. Since we use the derivative expansion of the vertices to compute the self-energies $\bar\Delta_\alpha^R$ (see Sec.~\ref{sec_rg}), the quasi-particle dispersion becomes linear to a very high degree of accuracy in the ``relativistic'' regime $|\p|\ll k_G$. In this regime, we expect the curvature of the dispersion to originate in the $(\p,\w)$ dependence of the self-energy $\bar\Delta_A^R(\p,\w)$ that is not included in the derivative expansion. Thus a reliable computation of the Beliaev damping would require a self-consistent numerical solution of the flow equations. 

While the Pad\'e approximant technique is very efficient to obtain $\bar\Delta_B^R(\p,\w)$, as well as $\bar\Delta^R_A(\p,\w)$ and $\bar\Delta_C^R(\p,\w)$ in the infrared regime, the computation of $\bar\Delta^R_A(\p,\w)$ and $\bar\Delta_C^R(\p,\w)$ for $|\p|\gg k_G$ appears more difficult for reasons that we do not fully understand. (Note also that the use of the derivative expansion might also be a source of difficulties for reasons discussed in Sec.~\ref{subsec_truncated}.) In the limit $|\p|\gg k_G$, the Bogoliubov approximation is however essentially correct and the corrections $\bar\Delta^R_A(\p,\w)$ and $\bar\Delta_C^R(\p,\w)$ provide a small broadening of the Bogoliubov quasi-particles (Beliaev damping) as can be directly verified from the one-loop self-energy diagrams.

\subsection{Spectral functions} 
\label{subsec_spectral}

\begin{figure}
\centerline{\includegraphics[width=6.5cm,clip]{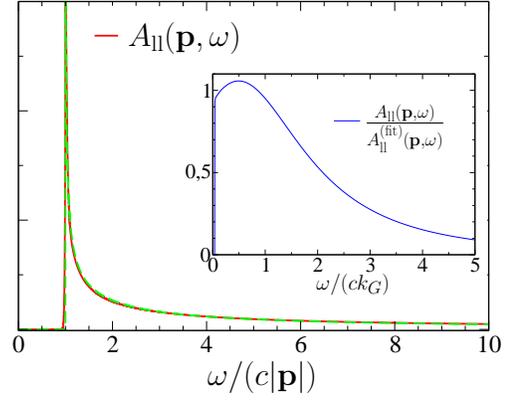}}
\caption{(Color online) Spectral function $\All(\p,\w)$ for $|\p|\simeq 0.036 k_G$ and $k=0$. The red solid line is the result obtained from the Pad\'e approximant, while the green dashed line corresponds to the analytic expression in (\ref{nr6}). The inset shows the ratio between $\All(\p,\w)$ and the approximate result (\ref{nr6}) on a larger energy scale.}
\label{fig_All}
\end{figure}

The knowledge of the retarded one-particle Green function enables to compute the spectral functions~\cite{note5}
\beq
\begin{split}
\All(\p,\w) &= - \frac{1}{\pi} \Im[\Gbarll^R(\p,\w)] , \\
\Att(\p,\w) &= - \frac{1}{\pi} \Im[\Gbartt^R(\p,\w)] , \\
\Alt(\p,\w) &= \frac{i}{\pi} \Re[\Gbarlt^R(\p,\w)] .
\end{split}
\label{nr5}
\eeq
From equations (\ref{propa_ir}) and (\ref{Gll}), we obtain 
\beq
\begin{split}
\Att(\p,\w) &= \frac{mcn_0}{\bar n|\p|} [\delta(\w-c|\p|) - \delta(\w+c|\p|)] , \\
\All(\p,\w) &= \frac{\sgn(\w)}{2\pi n_0C} \frac{\theta(|\w|-c|\p|)}{\sqrt{\w^2-(c\p)^2}} , \\
\Alt(\p,\w) &= i \frac{mc^2}{2\bar n} \frac{dn_0}{d\mu} [ \delta(\w-c|\p|) + \delta(\w+c|\p|)] ,
\end{split}
\label{nr6}
\eeq
in the infrared regime. $\Att(\p,\w)$ and $\Alt(\p,\w)$ exhibit Dirac peaks at the sound mode energy $\pm c|\p|$.
On the other hand, the longitudinal spectral function $\All(\p,\w)$ shows a critical continuum with a singularity at the Bogoliubov mode energy, in agreement with the predictions of the hydrodynamic approach~\cite{Giorgini92}. The spectral function $\All(\p,\w)$ obtained from the Pad\'e approximant is shown in Fig.~\ref{fig_All}. The square root singularity is very well reproduced and extends up to $\w\sim ck_G$. 

From these results, we can deduce the spectral function of the normal (U(1) invariant) Green function (see Appendix \ref{app_high_w}),  
\begin{align}
\An(\p,\w) &= -\frac{1}{\pi} \Im[\Gn^R(\p,\w)] \nonumber \\ &= \half \left[ \All(\p,\w) + \Att(\p,\w) \right] - i\Alt(\p,\w) \nonumber \\ 
&\simeq \half \left[ \All(\p,\w) + \Att(\p,\w) \right] .
\end{align}
The singularity of the longitudinal correlation function shows up as a continuum of excitations above the Bogoliubov sound mode. The respective spectral weights at positive frequencies of the transverse and longitudinal fluctuations are given by
\beq
\begin{split}
S_{\rm tt} &= \frac{mcn_0}{2\bar n|\p|} , \\ 
S_{\rm ll} &\simeq \frac{1}{4\pi n_0 C} \int_{c|\p|}^{ck_G} \frac{d\w}{\sqrt{\w^2-(c\p)^2}} \\
&\simeq \frac{1}{4\pi n_0C} \ln\left(\frac{2k_G}{|\p|}\right)
\end{split}
\eeq
for $|\p|\ll k_G$. Using the estimate (\ref{C_est}) of the constant $C$, we obtain the ratio
\beq
\begin{split}
\frac{S_{\rm ll}}{S_{\rm tt}} &\simeq \frac{mc|\p|}{8\pi\bar n} \ln\left(\frac{2k_G}{|\p|}\right) \\
& \simeq \frac{1}{8\pi} \left(\frac{mg}{\bar n}\right)^{1/2} |\p| \ln\left(\frac{2k_G}{|\p|}\right) ,
\end{split}
\eeq
where the last result is obtained with $c\simeq \sqrt{g\bar n/m}$. This ratio is extremely small in the weak coupling limit where $mg\ll 1$ and $|\p|\ll k_G\ll \sqrt{\bar n}$. It can however become sizable in the intermediate coupling regime when $mg\sim 1$ and $k_G$ is not much smaller than $\sqrt{\bar n}$. 

\section{Conclusion} 

The BMW NPRG method provides a powerful tool to study interacting boson systems. In particular, it enables to obtain the momentum and frequency dependence of the correlation functions on all energy scales. Our results reveal the crucial role of the Ginzburg scale $k_G$ in zero-temperature Bose superfluids. At large momenta or energies, $|\p|\gg k_G$ or $|\w|/c\gg k_G$, the Bogoliubov theory provides a good approximation to the correlation functions. For $|\p|,|\w|/c\ll k_G$, the correlation functions are governed by a different fixed point, which corresponds to Popov's hydrodynamic theory. Throughout the paper, we have emphasized that interacting boson systems can be understood within the framework of the (quantum) $O(2)$ model. The infrared behavior of this model is characterized by singular longitudinal fluctuations induced by the coupling to transverse (phase) fluctuations, a phenomenon which is common to all models with a continuous broken symmetry~\cite{Patasinskij73}.  

From a technical point, we have not solved the BMW equations in their full glory. By neglecting the field dependence of the self-energies $\Delta_\alpha(p;n)$ (which were approximated by $\Delta_\alpha(p;n_0)$) and using the derivative expansion, we have obtained flow equations which can be solved with reasonable numerical effort. Yet these equations yield a remarkable description of the singularity of the self-energy induced by the divergence of the longitudinal susceptibility. Quasi-particle life-time (Beliaev damping) can also be obtained in principle if the flow equations are solved self-consistently (\ie without relying on the derivative expansion).

We have restricted our analysis to the weak coupling limit where the two characteristic momentum scales $k_h$ and $k_G$ are well separated ($k_G\ll k_h\ll \bar n^{1/d}$). The characteristic momentum scale $\bar n^{1/d}$ does not play any role in this limit. When the dimensionless coupling constant is of order unity (intermediate coupling regime), the three characteristic scales become of the same order: $k_G\sim k_h\sim \bar n^{1/d}$. The momentum range $[k_G,k_h]$ where the linear spectrum can be described by the Bogoliubov theory is then suppressed. We expect the strong coupling regime to be governed by a single characteristic momentum scale, namely $\bar n^{1/d}$. A good description of physical phenomena at the scale of the interparticle spacing is likely to require the consideration of the complete BMW equations (with no additional approximation) with both the field and $(\p,\w)$ dependence of the vertices taken into account. 

In one dimension, superfluidity exists without Bose Einstein condensation ($n_0=0$), and our results regarding the infrared behavior of the correlation functions do not apply. If however, we insist on using the Bogoliubov theory as a starting point, we find from the perturbative estimate of Sec.~\ref{subsec_kG} a characteristic length $k_G \sim (gm)^{3/4}n_0^{1/4}$. This expression makes sense if we interpret $n_0$ as the condensate density $n_{0,k_G}$ at the scale $k_G$. A similar characteristic scale, $k_s\sim (gm)^{3/4}\bar n^{1/4}$, has been found in Ref.~\cite{Khodas08}. In weakly interacting one-dimensional Bose gases, $k_s$ separates a high-momentum regime ($|\p|>k_s$) where the Gross-Pitaevskii description is valid, from a low-momentum regime ($|\p|<k_s$) where a more elaborate description (e.g. based on the exact solution of the Lieb-Liniger model~\cite{Lieb63a,Lieb63b}) is required. The description of one-dimensional superfluidity from the NPRG is challenging, even if the derivative expansion yields reasonable results at weak coupling~\cite{Dupuis07}, and should be an interesting test of the BMW scheme.

\begin{acknowledgments}
I would like to thank B. Delamotte for enlightening discussions on the BMW scheme and its numerical implementation. I am also grateful to P. Kopietz for discussions, and to D. Gangardt for discussing the possible relevance of the Ginzburg scale $k_G$ in one-dimensional Bose gases and for pointing out Ref.~\cite{Khodas08}.
\end{acknowledgments} 

\appendix

\section{Bogoliubov's theory}
\label{app_bog}

In this Appendix, we briefly review the main results of Bogoliubov's theory. 

\subsection{Beliaev's self-energies} 

The action of interacting bosons is often written in terms of the two-component field
\beq
\Psi(p) = \left( \begin{array}{c} \psi(p) \\ \psi^*(-p)  \end{array} \right) , \quad 
\Psi^\dagger(p) = \bigl( \psi^*(p), \psi(-p) \bigr) ,
\eeq
where $p=(\p,i\w)$. The one-particle (connected) propagator then becomes a $2\times 2$ matrix whose inverse in Fourier space is given by
\beq
\left( 
\begin{array}{cc} i\w + \mu -\eps_\p -\Sign(p) & - \Sigan(p) \\
 -\Sigan^*(p) & -i\w + \mu -\eps_\p -\Sign(-p)
\end{array}
\right) ,
\label{propa}
\eeq
where $\Sign$ and $\Sigan$ are the normal and anomalous self-energies, respectively, and $\eps_\p=\p^2/2m$. Making use of (\ref{psidef}) and the relation $G=-\bar\Gamma^{(2)-1}$ between the propagator and the two-point vertex, one obtains equation (\ref{sigrel}) if one chooses a real order parameter $\mean{\psi(x)}=\sqrt{n_0}$. The normal and anomalous self-energies satisfy the Hugenholtz-Pines theorem~\cite{Hugenholtz59}
\beq
\Sign(0)-\Sigan(0) = \mu ,
\label{hpth} 
\eeq
which is a consequence of the spontaneously broken global U(1) symmetry in the superfluid phase. 

Using (\ref{propa}), we can relate the longitudinal propagator
\begin{align}
\Gll(p) &= - \mean{\psi_1(p)\psi_1(-p)}_c \nonumber \\ &= - \half \mean{ [\psi(p)+\psi^*(-p)][\psi(-p)+\psi^*(p)] }_c 
\end{align}
($\mean{\cdots}_c$ denotes for the connected part of the propagator)
to the self-energies $\Sign$ and $\Sigan$. Anticipating that $\mu-\Sign(p),\Sigan(p)\gg \p^2,\w^2$ when $\p,\w\to 0$ (and neglecting terms $\calO(\p^2,\w^2)$, we deduce  
\begin{align}
\lim_{p\to 0} \Gll(p) &= \lim_{p\to 0} \frac{\mu-\half[\Sign(p)+\Sign(-p)]+\Sigan(p)}{[\mu-\Sign(p)][\mu-\Sign(-p)]-\Sigan(p)^2} \nonumber \\ 
&= \lim_{p\to 0} \frac{-1}{2\Sigan(p)} \nonumber \\ 
&= \lim_{p\to 0} \frac{-1}{2n_0\bar\Gamma_B(p)},
\end{align} 
where we have used the Hugenholtz-Pines theorem (\ref{hpth}). 

\subsection{Bogoliubov's approximation} 
\label{subapp_ba}

The Bogoliubov approximation is based on the microscopic action (\ref{action4}) and a first-order computation of the self-energies
\beq
\begin{split}
\Sign^{\rm B}(p) &= 2gn_0 , \\
\Sigan^{\rm B}(p) &= gn_0 ,
\end{split}
\eeq
where the condensate density $n_0=\mu/g$. This yields the propagators 
\beq
\begin{split}
G_{\rm n}^{\rm B}(p) &= -\mean{\psi(p)\psi^*(p)}_c = \frac{-i\w-\eps_\p-gn_0}{\w^2+E_\p^2} ,\\ 
G_{\rm an}^{\rm B}(p) &= -\mean{\psi(p)\psi(-p)}_c = \frac{gn_0}{\w^2+E_\p^2} ,
\end{split}
\eeq
where $E_\p=[\eps_\p(\eps_\p+2gn_0)]^{1/2}$ is the Bogoliubov quasi-particle excitation energy. When $|\p|$ is larger than the healing momentum $k_h=(2gmn_0)^{1/2}$, the spectrum $E_\p\simeq \eps_\p+gn_0$ is particle-like, whereas it becomes sound-like for $|\p|\ll k_h$ with a velocity $c_B=\sqrt{gn_0/m}$. In the small-momentum limit $|\p|\ll k_h$,
\beq
\begin{split}
\GllB(p) &= - \frac{\eps_\p}{\w^2+c_B^2 \p^2} , \\ 
\GttB(p) &= - \frac{2gn_0}{\w^2+c_B^2 \p^2} , \\ 
\GltB(p) &= \frac{\w}{\w^2+c_B^2 \p^2} .
\end{split}
\eeq 
Note that in the Bogoliubov approximation, the occurrence of a linear spectrum is related to $\Sigan(0)$ being nonzero. In the weak coupling limit, $n_0$ is approximately given by the full density $\bar n$, and the healing momentum can also be defined by $k_h=(2gm\bar n)^{1/2}$ (which is the definition taken in Sec.~\ref{sec_nr}). 

\subsection{Perturbative estimate of the Ginzburg scale $k_G$}
\label{subsec_kG}

Let us consider the one-loop correction $\Sigan^{(1)}(p)$ to the Bogoliubov result $\Sigan^B(p)=gn_0$. The leading contribution comes from the one-loop diagram \newline
\centerline{
\includegraphics[width=3cm]{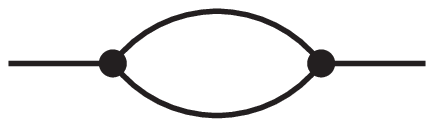}}
where the internal lines correspond to transverse fluctuations, i.e. 
\begin{align}
\Sigan^{(1)}(p) &\simeq - \frac{g^2n_0}{2} \int_q \Gtt(q)\Gtt(p+q) \nonumber \\ 
& \simeq -\frac{g^2n_0}{2} \frac{S_d}{(2\pi)^d} \int_k^{k_h} d|\q||\q|^{d-1} \nonumber \\ & \hspace{2cm} \times \intinf \frac{d\w}{2\pi} \left( \frac{2gn_0}{\w^2+c_B^2\q^2}\right)^2 \nonumber \\ 
& \simeq -\frac{g}{2} (gmn_0)^{3/2} \frac{S_d}{(2\pi)^d} \int_k^{k_h} \frac{d|\q|}{|\q|^{4-d}} ,
\end{align}
where $k_h$ is the healing momentum defined in section \ref{subapp_ba}, and $S_d$ the surface of the unit sphere in $d$ dimensions. The infrared limit $k$ in the integral is of order $(\p^2+\wnu^2/c_B^2)^{1/2}$ (with $p=(\p,i\wnu)$). The one-loop correction is divergent when $d\leq 3$. This divergence reflects the difficulties of diagrammatic calculations beyond the Bogoliubov approximation and is a manifestation of the diverging longitudinal susceptibility~\cite{Weichman88}. We estimate the Ginzburg momentum scale $k_G$ from the condition $|\Sigan^{(1)}(p)| \sim |\Sigan^B(p)|$ [see Eq.~(\ref{kg_est})].

\section{Symmetries and Ward identities} 
\label{app_wi} 

\subsection{Gauge invariance}

Let us consider the microscopic action 
\begin{align}
S = \int dx & \Bigl[ \psi^*(x)\Bigl(\dtau-\mu(x) - \frac{1}{2m}[\nablabf-i\A(x)]^2
  \Bigr) \psi(x) \nonumber \\ & + \frac{g}{2} |\psi(x)|^4 \Bigr]
\label{action5}
\end{align}
in the presence of external sources $\mu(x)$ and $\A(x)$. $S$ is invariant in the gauge transformation
\beq
\begin{split} \\
\psi(x) &\to \psi(x) e^{i\alpha(x)} , \\
\psi^*(x) &\to \psi^*(x) e^{-i\alpha(x)} , \\
\mu(x) &\to \mu(x) + i \dtau \alpha(x) \\ 
\A(x) &\to \A(x) + \nablabf\alpha(x) ,
\end{split}
\eeq
where $\alpha(x)$ is an arbitrary real function. This implies that the effective action satisfies
\beq
\Gamma[R(\alpha)\phi,\mu+i\dtau\alpha,A_\nu+\partial_\nu \alpha]=\Gamma[\phi,\mu,A_\nu] ,
\label{gauge1}
\eeq
where $\phi=(\phi_1,\phi_2)^T$ and 
\beq 
R(\alpha) = \left( 
\begin{array}{cc} \cos(\alpha) & -\sin(\alpha) \\ \sin(\alpha) & \cos(\alpha) \end{array} \right) 
\eeq
is a two-dimensional rotation matrix. Differentiating (\ref{gauge1}) with respect to $\alpha(x)$, we obtain
\beq
\sum_{i,j} \frac{\delta\Gamma}{\delta\phi_i(x)} \eps_{ij} \phi_j(x) + i\dtau \frac{\delta\Gamma}{\delta\mu(x)} + \sum_\nu \partial_\nu \frac{\delta\Gamma}{\delta A_\nu(x)} = 0 . 
\eeq
Differentiating now with respect to $\phi_l(x_2)$ and $\mu(x_2)$ and setting $\phi=(\sqrt{2n_0},0)$, $\mu(x)=\mu$ and $\A(x)=0$ gives  
\beq
\begin{split}
-\sqrt{2n_0} \bar\Gamma^{(2)}_{l2}(x_2,x_1) & + i\partial_{\tau_1} \bar\Gamma^{(2)}_{l;0}(x_2,x_1) \\ & + \sum_{\nu_1} \partial_{\nu_1} \bar\Gamma^{(2)}_{l;\nu_1}(x_2,x_1) = 0 , \\ 
-\sqrt{2n_0} \bar\Gamma^{(2)}_{2;0}(x_1,x_2) & + i\partial_{\tau_1} \bar\Gamma^{(2)}_{;00}(x_1,x_2) \\  & + \sum_{\nu_1} \partial_{\nu_1} \bar\Gamma^{(2)}_{;\nu_10}(x_1,x_2) = 0 ,
\end{split}
\label{gauge2}
\eeq
where we have introduced
\beq
\begin{split}
\Gamma^{(2)}_{l;0}(x_2,x_1) &= \frac{\delta^{(2)}\Gamma}{\delta\phi_l(x_2) \delta\mu(x_1)} , \\
\Gamma^{(2)}_{;00}(x_2,x_1) &= \frac{\delta^{(2)}\Gamma}{\delta\mu(x_2) \delta\mu(x_1)}, \end{split}
\eeq
and similar definitions for $\Gamma^{(2)}_{l;\nu}(x_2,x_1)$ and $\Gamma^{(2)}_{;\nu0}(x_2,x_1)$. Note that with the choice $\phi=(\sqrt{2n_0},0)$, we can identify $\bar\Gamma_{12}^{(2)}$ to $\bar\Gamma_C$, and $\bar\Gamma_{22}^{(2)}$ to $\bar\Gamma_A$. In Fourier space, (\ref{gauge2}) leads to the Ward identities 
\begin{align}
\sqrt{2n_0} \bar\Gamma^{(2)}_{12}(p) + \w \bar\Gamma^{(2)}_{1;0}(p) + \sum_\nu ip_\nu \bar\Gamma^{(2)}_{1;\nu}(p) &= 0 , \label{wi1} \\
\sqrt{2n_0} \bar\Gamma^{(2)}_{22}(p) + \w \bar\Gamma^{(2)}_{2;0}(p) + \sum_\nu ip_\nu \bar\Gamma^{(2)}_{2;\nu}(p) &= 0 , \label{wi2} \\ 
\sqrt{2n_0} \bar\Gamma^{(2)}_{2;0}(p) - \w \bar\Gamma^{(2)}_{;00}(p) - \sum_\nu ip_\nu \bar\Gamma^{(2)}_{;\nu 0}(p) &= 0 . \label{wi3}  
\end{align}
From (\ref{wi1}), we deduce
\begin{align}
\frac{\partial}{\partial\w} \bar\Gamma^{(2)}_{12}(p)\Bigl|_{p=0} &= - \frac{1}{\sqrt{2n_0}} \bar\Gamma^{(2)}_{1;0}(p=0) , \nonumber \\ 
&= - \frac{1}{\sqrt{2n_0}} \frac{\partial^2 U}{\partial\phi_1\partial\mu}\biggl|_{n_0} = - \frac{\partial^2 U}{\partial n\partial\mu}\biggl|_{n_0} ,
\end{align}
where the effective potential $U(n,\mu)$ is considered as a function of both $n$ and $\mu$. From (\ref{wi2}) and (\ref{wi3}), we obtain
\beq
\begin{split}
\frac{\partial}{\partial\w^2} \bar\Gamma^{(2)}_{22}(p)\Bigl|_{p=0} &=  - \frac{1}{\sqrt{2n_0}} \frac{\partial}{\partial\w} \bar\Gamma^{(2)}_{2;0}(p)\Bigl|_{p=0} , \\ 
\frac{\partial}{\partial\w} \bar\Gamma^{(2)}_{2;0}(p)\Bigl|_{p=0} &=  \frac{1}{\sqrt{2n_0}} \bar\Gamma^{(2)}_{;00}(0) = \frac{1}{\sqrt{2n_0}} \frac{\partial^2 U}{\partial\mu^2}\biggl|_{n_0} 
\end{split}
\eeq
and therefore 
\beq
\frac{\partial}{\partial\w^2} \bar\Gamma^{(2)}_{22}(p)\Bigl|_{p=0} = - \frac{1}{2n_0} \frac{\partial^2 U}{\partial\mu^2}\biggl|_{n_0} .
\eeq

\subsection{Galilean invariance}

Another Ward identity can be obtained from the Galilean invariance of the microscopic action. The latter is invariant in the transformation $\psi(x)\to \psi(x)e^{i\q\cdot\r}$, $\psi^*(x)\to \psi(x)e^{-i\q\cdot\r}$ if we shift the chemical potential $\mu$ by $\q^2/2m$, which implies
\beq
\Gamma[R(\alpha)\phi,\mu+\q^2/2m] = \Gamma[\phi,\mu],
\label{gauge3}
\eeq
where $\alpha(x)=\q\cdot\r$ and the chemical potential $\mu$ is taken uniform and time independent. To order $\q^2$, equation (\ref{gauge3}) gives
\beq
\begin{split}
0 &= \frac{\q^2}{2m} \frac{\partial\bar\Gamma}{\partial\mu} + n_0 \int dxdx' \bar\Gamma^{(2)}_{22}(x-x') \alpha(x) \alpha(x') \\
&= \frac{\q^2}{2m} \frac{\partial\bar\Gamma}{\partial\mu} + \beta V n_0 \q^2 \frac{\partial}{\partial \p^2} \bar\Gamma^{(2)}_{22}(p)\Bigl|_{p=0} ,
\end{split}
\eeq
where we have set $\phi=(\sqrt{2n_0},0)$. Since 
\beq
\frac{\partial\bar\Gamma}{\partial\mu} = \beta V \frac{\partial U}{\partial\mu}\biggl|_{n_0} = - \beta V \bar n
\eeq
(see Eq.~(\ref{nbar})), we finally obtain
\beq
\frac{\partial}{\partial \p^2} \bar\Gamma^{(2)}_{22}(p)\Bigl|_{p=0} = \frac{\bar n}{2mn_0} ,
\eeq
where $\bar n$ is the mean boson density.

\begin{widetext}

\section{Flow equations}
\label{app_floweq} 

\subsection{BMW equations} 

In the BMW approximation, the flow equation of the two-point vertex is given by
\begin{multline}
\dt \Gamma^{(2)}_{ij}(p;\phi) = -\half \sum_{q,i_1,i_2} \tilde\dt G_{i_1i_2}(q;\phi) \Gamma^{(4)}_{iji_2i_1}(p,-p,0,0;\phi)  \\ 
- \half \sum_{q,i_1\cdots i_4} \Bigl\lbrace \Gamma^{(3)}_{ii_2i_3}(p,0,-p;\phi) \Gamma^{(3)}_{ji_4i_1}(-p,p,0;\phi) % \\ \times
[\tilde\dt G_{i_1i_2}(q;\phi)]G_{i_3i_4}(p+q;\phi) + (p\leftrightarrow -p, i\leftrightarrow j) \Bigr\rbrace ,
\label{flow2}
\end{multline}
where the three- and four-point vertices in (\ref{flow2}) are obtained from the field derivatives of the two-point vertex [Eq.~(\ref{gam_bmw})]. From (\ref{sym1}) and (\ref{Gdef}), we obtain 
\beq
\begin{split}
G_{ij}(p;\phi) ={}& \delta_{i,j}[\delta_{i,1} \Gll(p;n) + \delta_{i,2} \Gtt(p;n) ] + \eps_{ij} \Glt(p;n) , \\ 
\Gamma^{(2)}_{ij}(p;\phi) ={}& \delta_{i,j}[ \Gamma_A(p;n) + \delta_{i,1} 2n \Gamma_B(p;n) ] + \eps_{ij} \Gamma_C(p;n) , \\  
\Gamma^{(3)}_{ijl}(p,-p,0;\phi) ={}& \frac{\sqrt{2n}}{\sqrt{\beta V}} \Bigl\lbrace 
\delta_{i,j}\delta_{l,1} \bigl[\Gamma_A'(p;n)+\delta_{i,1}2n\Gamma_B'(p;n)\bigr] + 
(\delta_{i,l}\delta_{j,1} + \delta_{j,l}\delta_{i,1}) \Gamma_B(p;n) + \eps_{ij} \delta_{l,1} \Gamma_C'(p;n) \Bigr\rbrace,  \\ 
\Gamma^{(4)}_{ijlm}(p,-p,0,0;\phi) ={}& \frac{1}{\beta V} \Bigl\lbrace \delta_{i,j}\delta_{l,m} \bigl[\Gamma'_A(p;n) + \delta_{l,1} 2n \Gamma''_A(p;n) \bigr] 
+ ( \delta_{i,l}\delta_{j,m} + \delta_{j,l}\delta_{i,m} ) \Gamma_B(p;n) \\ 
& + \bigl[ \delta_{i,1} (\delta_{j,m} \delta_{l,1} + \delta_{l,m} \delta_{j,1}) 
+ \delta_{m,1} (\delta_{i,l} \delta_{j,1} + \delta_{j,l} \delta_{i,1}) + \delta_{i,m}\delta_{j,1}\delta_{l,1} \bigr] 2n \Gamma'_B(p;n) \\ &
+ \delta_{i,1} \delta_{j,1} \delta_{l,1} \delta_{m,1} 4n^2 \Gamma''_B(p;n) 
+ \delta_{l,m}\eps_{ij} \bigl[\Gamma'_C(p;n)  + \delta_{l,1} 2n \Gamma''_C(p;n) \bigr] \Bigl\rbrace 
\end{split}
\eeq
($\Gamma_\alpha'(p;n) = \partial_n \Gamma_\alpha(p;n)$, etc.) 
for the particular field configuration $\phi=(\sqrt{2n},0)$. The flow equation (\ref{flow2}) then gives
\begin{align} 
\dt \Gamma_A(p;n) ={}& - \half \Ill(n) [ \Gamma'_A(p;n)+2n\Gamma''_A(p;n)] 
                       - \half \Itt(n) [ \Gamma'_A(p;n)+2\Gamma_B(p;n)] \nonumber \\ &
-2n \bigl[ \Jlltt(p;n) \Gamma_A'(p;n)^2 + \Jttll(p;n) \Gamma_B(p;n)^2 + 2 \Jltlt(p;n) \Gamma'_A(p;n) \Gamma_B(p;n) \nonumber \\ & 
- \Jllll(p;n) \Gamma'_C(p;n)^2 - 2\Jlllt(p;n) \Gamma_A'(p;n) \Gamma'_C(p;n) + 2 \Jltll(p;n) \Gamma_B(p;n) \Gamma'_C(p;n) \bigl] , \label{eq1_bmw} \\
\dt \Gamma_B(p;n) ={}& \frac{1}{2n} [\Itt(n) - \Ill(n)] \Gamma_B(p;n) - \Ill(n) \left[ \frac{5}{2} \Gamma'_B(p;n) + n\Gamma''_B(p;n) \right] - \half \Itt(n) \Gamma'_B(p;n) 
\nonumber \\ &
- \Jllll(p;n) \bigl[X(n)^2 + \Gamma'_C(p;n)^2 \bigr] + \Jlltt(p;n) \bigl[ \Gamma'_A(p;n)^2 + \Gamma'_C(p;n)^2 \bigr] 
+ \Jttll(p;n) \Gamma_B(p;n)^2 \nonumber \\ & - \Jtttt(p;n) \Gamma_B(p;n)^2 
+ 2 \Jltlt(p;n)\Gamma_B(p;n) \bigl[X(n) + \Gamma'_A(p;n) \bigr] + 2 \Jlllt(p;n) \bigl[ X(n)-\Gamma'_A(p;n) \bigr] \Gamma'_C(p;n) \nonumber \\ &
+ 2 \Jltll(p;n) \Gamma_B(p;n) \Gamma'_C(p;n) + 2 \Jlttt(p;n) \Gamma_B(p;n) \Gamma'_C(p;n), \label{eq2_bmw} \\ 
\dt \Gamma_C(p;n) ={}& - \half \Ill(n) \bigl[ \Gamma'_C(p;n)+2n\Gamma''_C(p;n) \bigr] - \half \Itt(n) \Gamma'_c(p;n) \nonumber \\ &
-2n \Bigl\lbrace \Jllll(p;n) X(n) \Gamma'_C(p;n) + \Jlltt(p;n) \Gamma'_A(p;n) \Gamma'_C(p;n) - \Jttlt(p;n) \Gamma_B(p;n)^2  \nonumber \\ &
+ \Jltll(p;n) \Gamma_B(p;n) \Gamma'_A(p;n) + \Jlllt(p;n) \bigl[ X(n) \Gamma'_A(p;n) - \Gamma'_C(p;n)^2 \bigr] - \Jltll(p;n) X(n) \Gamma_B(p;n) \Bigr\rbrace , 
\label{eq3_bmw}
\end{align}
\end{widetext}
where $X=\Gamma'_A+2\Gamma_B+2n\Gamma'_B$. The coefficients $I_\alpha(n)$ and $J_{\alpha\beta}(n;p)$ are defined in (\ref{IJdef}). If we set $\Gamma_C=0$ and $p=(\p,0)$, we reproduce the flow equations of the classical O(2) model derived in Ref.~\cite{Benitez08}. 

\subsection{Truncated flow equations}

The flow equations simplify considerably when the field dependence of the self-energy $\Delta_\alpha(p;n)$ is neglected and the effective potential $U(n)$ expanded about $n_0$ as in (\ref{Utrunc}). In this case the only non-vanishing field derivative is $\Gamma'_A(p;n)=\lambda$ while $\Gamma'_B(p;n)=\Gamma'_C(p;n)=0$ [see Eqs.~(\ref{deltadef})], so that we obtain
\beq
\begin{split}
\dt \Gamma_A( p;n) ={}& -\half \Ill(n) \lambda - \half \Itt(n)[\lambda + 2\Gamma_B(p;n)] \\ &
-2n \bigl[ \Jlltt(p;n) \lambda^2 + \Jttll(p;n) \Gamma_B^2(p;n) \\ & +2 \Jltlt(p;n) \lambda \Gamma_B(p;n) \bigr] , \\ 
\dt \Gamma_B( p;n) ={}& \frac{1}{2n} [\Itt(n)-\Ill(n)] \Gamma_B(p;n) + \Jlltt(p;n)\lambda^2 \\ &
- \Jllll(p;n)[\lambda+2\Gamma_B(p;n)]^2 \\ & 
+ [\Jttll(p;n) - \Jtttt(p;n)] \Gamma_B^2(p;n) \\ &
+ 4 \Jltlt(p;n) \Gamma_B(p;n) [ \lambda+\Gamma_B(p;n) ] , \\
\dt \Gamma_C( p;n) ={}& 2n \bigl\lbrace \Jttlt(p;n) \Gamma^2_B(p;n) \\ & 
- \Jlttt(p;n) \lambda \Gamma_B(p;n) \\ &
- \Jlllt(p;n) \lambda [\lambda+2\Gamma_B(p;n)] \\ &
+ \Jltll(p;n) \Gamma_B(p;n) [\lambda+2\Gamma_B(p;n)] \bigr\rbrace .
\end{split}
\label{gammaflow}
\eeq

We can finally deduce the flow equations for the self-energy $\bar\Delta_\alpha(p)=\Delta_\alpha(p;n_0)$ from its definition (\ref{deltadef}), 
\beq
\begin{split} 
\dt \bar\Delta_A(p) &= \dt \Gamma_A(p;n)|_{n_0} + \Gamma_A'(p;n_0) \dt n_0 , \\
\dt \bar\Delta_B(p) &= \dt \Gamma_B(p;n)|_{n_0} + \Gamma_B'(p;n_0) \dt n_0 -\dt\lambda,\\
\dt \bar\Delta_C(p) &= \dt \Gamma_C(p;n)|_{n_0} + \Gamma_C'(p;n_0) \dt n_0 , 
\end{split}
\label{deltaflow}
\eeq
where 
\beq
\begin{split}
\Gamma_A'(p;n_0) &= U''(n_0)=\lambda, \\ 
\Gamma_B'(p;n_0) &= U'''(n_0)=0, \\
\Gamma_C'(p;n_0) &= 0 .
\end{split}
\eeq
This leads to equations (\ref{flow3}) and (\ref{flow6}). 

\subsection{Dimensionless flow equations} 
\label{app_dimensionless}

For numerically solving the flow equations, it is useful to introduce the dimensionless variables (\ref{dimless_def}) as well as the dimensionless self-energy, 
\beq
\begin{split}
\tilde \Delta_A(p) &= (Z_A\eps_k)^{-1} \bar\Delta_A(p) ,\\
\tilde \Delta_B(p) &= (k^{-d}\eps_k Z_AZ_C)^{-1} \bar\Delta_B(p) , \\
\tilde \Delta_C(p) &= (Z_A\eps_k)^{-1} \bar\Delta_C(p) ,
\end{split} 
\eeq
where $\eps_k = k^2/2m$. In dimensionless form, equations (\ref{flow5},\ref{flow3},\ref{flow6}) read
\beq
\begin{split}
\dt \tilde n_0 ={}& -(d+\eta_C) \tilde n_0 + \frac{3}{2} \Itildell + \half \Itildett , \\
\dt \tilde \lambda ={}& (d-2+\eta_A+\eta_C) \tilde\lambda \\ & - \tilde\lambda^2 \bigl[9\Jtildellll(0) - 6\Jtildeltlt(0) + \Jtildetttt(0) \bigr] , \\ 
\eta_A ={}&  2\tilde n_0\tilde \lambda^2 \frac{\partial}{\partial y} \bigl[\Jtildelltt(p)+\Jtildettll(p)+2\Jtildeltlt(p)\bigr]_{p=0} , \\ 
\eta_C ={}& -2 \tilde n_0\tilde\lambda^2 \frac{\partial}{\partial\tilde\w} \bigl[\Jtildettlt(p) - \Jtildelttt(p)  \\ & -3\Jtildelllt(p) + 3\Jtildeltll(p) \bigr]_{p=0} , \\ 
\dt \tilde V_A ={}& (2-\eta_A+2\eta_C)\tilde V_A \\ & -2 \tilde n_0\tilde\lambda^2 \frac{\partial}{\partial\tilde\w^2} \bigl[\Jtildelltt(p)+\Jtildettll(p)+2\Jtildeltlt(p)\bigr]_{p=0} ,
\end{split}
\label{flow_dim}
\eeq
and 
\begin{align}
\dt \tilde\Delta_A(p) ={}& (\eta_A-2)\tilde\Delta_A(p) + \tilde\lambda(\Itildell-\Itildett) \nonumber \\ & -2\tilde n_0\tilde\lambda^2  \bigl[\Jtildelltt(p)+\Jtildettll(p)+2\Jtildeltlt(p)\bigr] , \nonumber \\
\dt \tilde\Delta_B(p) ={}&  (d-2+\eta_A+\eta_C) \tilde\Delta_B(p) + \frac{\tilde\lambda}{2\tilde n_0}(\Itildett-\Itildell) \nonumber\\ & +\tilde\lambda^2 \bigl[-9\Jtildellll(p)+\Jtildelltt(p) \nonumber \\ & +\Jtildettll(p)-\Jtildetttt(p)+8\Jtildeltlt(p) \bigr] \\ \nonumber &
+ \tilde\lambda^2 \bigl[ 9\Jtildellll(0)- 6 \Jtildeltlt(0) + \Jtildetttt(0) \bigr] , \\ 
\dt \tilde\Delta_C(p) ={}& (\eta_A-2)\tilde\Delta_C(p) + 2 n_0\lambda^2 \bigl[\Jtildettlt(p) - \Jtildelttt(p)\nonumber  \\ & -3\Jtildelllt(p) + 3\Jtildeltll(p) \bigr] , \nonumber 
\end{align} 
where 
\beq
y = \frac{\p^2}{k^2}, \qquad \tilde\w = \frac{Z_C}{Z_A\eps_k} \w ,
\label{y_def}
\eeq
and $\eta_A=-\dt\ln Z_A$, $\eta_C=-\dt\ln Z_C$. The coefficients $\tilde I_\alpha$ and $\tilde J_{\alpha\beta}(p)$ are defined in section \ref{IJ_dim}.

\subsection{Coefficients $I_\alpha$ and $J_{\alpha\beta}(p)$}

\subsubsection{$\bar I_\alpha$ and $\bar J_{\alpha\beta}(p)$}

To compute the coefficients $\bar I_\alpha=I_\alpha(n_0)$ and $\bar J_{\alpha\beta}(p) = J_{\alpha\beta}(p;n_0)$ and their derivatives with respect to $\p$ or $\w$, it is convenient to introduce 
\beq
\begin{split}
A(p) &= \bar\Gamma_A(p) + R(p) , \\ 
B(p) &= A(p) + 2n_0\bar\Gamma_B(p), \\ 
C(p) &= \bar\Gamma_C(p), \\ 
D(p) &= C(p)^2 + A(p)B(p) . 
\end{split}
\eeq
With these notations, we have
\beq
\Gbarll = - \frac{A}{D} , \quad
\Gbartt = - \frac{B}{D} , \quad
\Gbarlt = \frac{C}{D} , 
\label{B4}
\eeq
and 
\beq
\begin{split} 
\tilde \dt \Gbarll &= - \dot R \frac{C^2-A^2}{D^2} , \\ 
\tilde \dt \Gbartt &= - \dot R \frac{C^2-B^2}{D^2} , \\ 
\tilde \dt \Gbarlt &= - \dot R \frac{C(A+B)}{D^2} , 
\end{split}
\label{B5}
\eeq
where
\beq
\begin{split}
\dot R &= -Z_A \eps_k Y (\eta_A r+2Yr') , \\
Y &= \frac{\p^2}{k^2} + \frac{\w^2}{c_0^2 k^2} ,
\end{split} 
\eeq
with $r\equiv r(Y)$ and $r'=\partial r/\partial Y$. Equations (\ref{B4}) and (\ref{B5}) can be used to compute $\bar I_\alpha$ and $\bar J_{\alpha\beta}(p)$, as well as $\partial_\w \bar J_{\alpha\beta}(p)|_{\w=0}$ and $\partial_{\w^2} \bar J_{\alpha\beta}(p)|_{\w=0}$.

\subsubsection{$\partial_{\eps_\p} \bar J_{\alpha\beta}(p)|_{p=0}$} 

Using
\begin{align}
\bar G_\alpha(p+q) ={}& \bar G_\alpha(q) + (\p^2+2\p\cdot\q) \bar G'_\alpha(q) \nonumber \\ &
+ 2(\p\cdot\q)^2 \bar G''_\alpha(q) + \calO(|\p|^3), 
\end{align}
for $p=(\p,0)$, we find 
\beq
\bar J_{\alpha\beta}(p) = \int_q \bigl(\tilde\dt \bar G_\alpha\bigr) \Bigl( \bar G_\beta + \p^2 \bar G'_\beta  + \frac{2}{d} \p^2 \q^2\bar G''_\beta \Bigr) + \calO(|\p|^4), 
\eeq
and 
\begin{align}
\frac{\partial}{\partial\p^2} \bar J_{\alpha\beta}(p) \Bigl|_{p=0} ={}& 4v_d \int_\w \int_0^\infty d|\q| |\q|^{d-1} \bigl(\tilde\dt \bar G_\alpha\bigr) \nonumber \\ & \times \Bigl(\bar G'_\beta + \frac{2}{d} \q^2 \bar G''_\beta \Bigl) , 
\label{Jder1}
\end{align} 
where we use the notation 
\beq
\bar G'_\alpha = \frac{\partial}{\partial \q^2} \bar G_\alpha
\eeq
(note that $\bar G_\alpha(p)$ is function of $\p^2$). We have introduced $v_d=[2^{d+1}\pi^{d/2}\Gamma(d/2)]^{-1}$. Using the variable $x=\q^2$ and integrating the last term of (\ref{Jder1}) by part, we find
\beq
\frac{\partial}{\partial\p^2} \bar J_{\alpha\beta}(p) \Bigl|_{p=0} = -8 \frac{v_d}{d} \int_\w \int_0^\infty d|\q|\,  |\q|^{d+1} \bigl(\tilde\dt \bar G'_\alpha\bigr) \bar G'_\beta .
\eeq
The operator $\tilde\dt$ is defined by 
\beq
\tilde\dt = \dot R \frac{\partial}{\partial R} + \dot R' \frac{\partial}{\partial R'} ,
\eeq
where 
\beq
\begin{split}
R' &= \frac{Z_A}{2m} (r+Yr'), \\
\dot R' &= - \frac{Z_A}{2m} [\eta_A r+ (\eta_A+4)Yr' + 2Y^2r''] .
\end{split}
\eeq
This gives 
\begin{multline}
\frac{\partial}{\partial\eps_\p} \bar J_{\alpha\beta}(p) \Bigl|_{p=0} = 4 \frac{v_d}{d} k^{d+2} Z_A  \int_\w  \int_0^\infty dy\, y^{d/2} \\ \times \Bigl\lbrace k^2 Y (\eta_A r+2Yr') \frac{\partial}{\partial R} \bar G'_\alpha \\
+ [\eta_Ar + (\eta_A+4)Yr' + 2Y^2r''] \frac{\partial}{\partial R'} \bar G'_\alpha \Bigr\rbrace \bar G'_\beta  . 
\label{B6}
\end{multline}
The function $\bar G'_\alpha(q)$ can be expressed as  
\beq
\begin{split}
\Gbarll' &= - \frac{1}{D^2} \bigl(C^2A'-A^2 B' \bigr) , \\
\Gbartt' &= - \frac{1}{D^2} \bigl(C^2B'-A' B^2 \bigr) , \\
\Gbarlt' &= - \frac{C}{D^2} (A'B + AB') , 
\end{split}
\label{B1}
\eeq 
where $A' = \partial_{\q^2} A$ and $B' = \partial_{\q^2} B$. Using 
\beq
\begin{split} 
\frac{\partial}{\partial R} &= \frac{\partial}{\partial A} + \frac{\partial}{\partial B}, \\ 
\frac{\partial}{\partial R'} &= \frac{\partial}{\partial A'} + \frac{\partial}{\partial B'},
\end{split}
\eeq
we obtain
\beq
\begin{split}
\frac{\partial}{\partial R} \Gbarll' ={}& \frac{2}{D^3} \bigl[ C^2(AB'+A'B  +AA') - A^3B' \bigr] , \\ 
\frac{\partial}{\partial R} \Gbartt' ={}& \frac{2}{D^3} \bigl[ C^2(AB'+A'B  +BB') - A'B^3 \bigr] , \\ 
\frac{\partial}{\partial R} \Gbarlt' ={}& -\frac{C}{D^3} \bigl[ (C^2-AB) (A'+B') \\ & - 2 A^2B' - 2 A'B^2 \bigr] ,
\end{split}
\label{B2}
\eeq
and
\beq
\begin{split}
\frac{\partial}{\partial R'} \Gbarll' ={}& -\frac{1}{D^2} \bigl(C^2 - A^2 \bigr) , \\ 
\frac{\partial}{\partial R'} \Gbartt' ={}& -\frac{1}{D^2} \bigl(C^2 - B^2 \bigr) , \\ 
\frac{\partial}{\partial R'} \Gbarlt' ={}& -\frac{C}{D^2} (A + B) . 
\end{split}
\label{B3}
\eeq
Equations (\ref{B1}), (\ref{B2}) and (\ref{B3}) are used to compute $\partial_{\eps_\p} \bar J_{\alpha\beta}(p)|_{p=0}$. In the derivative expansion, we use the simplified expressions
\beq
\begin{split}
A(p) &= V_A\w^2 + Z_A\eps_\p + R(p) , \\ 
B(p) &= A(p) + 2n_0\lambda, \\ 
C(p) &= Z_C\w, 
\end{split}
\label{B7}
\eeq
and
\beq
A' = B' = \frac{Z_A}{2m}(1+r+Yr') , 
\label{B8}
\eeq

\subsubsection{$\tilde I_\alpha$ and $\tilde J_{\alpha\beta}(p)$}
\label{IJ_dim} 

We introduce dimensionless propagators,
\beq
\begin{split}
\Gtildell &= \frac{\Gll}{Z_A\eps_k} = - \frac{\tilde A}{\tilde D} , \\ 
\Gtildett &= \frac{\Gtt}{Z_A\eps_k} = - \frac{\tilde B}{\tilde D} , \\ 
\Gtildelt &= \frac{\Glt}{Z_A\eps_k} = \frac{\tilde C}{\tilde D} ,
\end{split}
\eeq
where
\beq
\begin{split}
\tilde A &= (Z_A\eps_k)^{-1} A , \\ 
\tilde B &= (Z_A\eps_k)^{-1} B , \\ 
\tilde C &= (Z_A\eps_k)^{-1} C .
\end{split}
\eeq
The dimensionless coefficients $\tilde I_\alpha$ and $\tilde J_{\alpha\beta}(p)$ are then defined by
\beq
\begin{split}
\tilde I_\alpha &= k^{-d} Z_C \bar I_\alpha \\ &= -2v_d \int_{y,\tilde\w} y^{d/2-1}(\eta_Ar+2Yr') \frac{\partial \tilde G_\alpha}{\partial r} 
\end{split} 
\eeq
and
\beq
\begin{split}
\tilde J_{\alpha\beta}(p) ={}& \frac{Z_AZ_C\eps_k}{k^d} \bar J_{\alpha\beta}(p) \\ 
={}& - \frac{1}{8\pi^2} \int_0^{(4-d)\pi} d\theta\sin^{d-2}\theta \\ &\times \int_{y,\tilde\w}  y^{d/2-1}(\eta_Ar+2Yr') \frac{\partial \tilde G_\alpha(q)}{\partial r} \tilde G_\beta(p+q) , \\ 
\tilde J_{\alpha\beta}(0) ={}& -2v_d \int_{y,\tilde\w}  y^{d/2-1}(\eta_Ar+2Yr') \frac{\partial \tilde G_\alpha}{\partial r} \tilde G_\beta 
\end{split}
\label{B9}
\eeq
($d=3$ or $d=2$). $y$ and $\tilde\w$ are defined in (\ref{y_def}). To compute (\ref{B9}), we use 
\beq
\begin{split} 
\frac{\partial \Gtildell}{\partial r} &= \frac{Y}{\tilde D^2} (\tilde A^2-\tilde C^2) , \\ 
\frac{\partial \Gtildett}{\partial r} &= \frac{Y}{\tilde D^2} (\tilde B^2-\tilde C^2) , \\ 
\frac{\partial \Gtildelt}{\partial r} &= -\frac{Y\tilde C}{\tilde D^2} (\tilde A+\tilde B) .
\end{split}
\eeq
In dimensionless form, Eq.~(\ref{B6}) becomes 
\begin{align}
\frac{\partial}{\partial y} \tilde J_{\alpha\beta}(p)\Bigl|_{p=0} ={}& \frac{Z_AZ_C\eps_k^2}{k^d} \frac{\partial}{\partial \eps_\p} J_{\alpha\beta}(p)\Bigl|_{p=0} \nonumber \\ 
={}& 4\frac{v_d}{d} \int_{y,\tilde\w} y^{d/2} \Bigl\lbrace (\eta_Ar+2Yr') \frac{\partial \tilde G'_\alpha}{\partial r}\biggl|_{r+Yr'} \nonumber \\ &
+\left[ \eta_Ar+(\eta_A+4)Yr' +2Y^2r'' \right] \nonumber \\ & \times Y^{-1} \frac{\partial \tilde G'_\alpha}{\partial r'} \Bigr\rbrace \tilde G'_\beta , 
\label{B11}
\end{align}
where ($\tilde G'_\alpha = \partial_y \tilde G_\alpha$)
\beq
\begin{split} 
\Gtildell' &= - \frac{1}{\tilde D^2} (\tilde C^2\tilde A' - \tilde A^2\tilde B') , \\ 
\Gtildett' &= - \frac{1}{\tilde D^2} (\tilde C^2\tilde B' - \tilde B^2\tilde A') , \\ 
\Gtildelt' &= - \frac{\tilde C}{\tilde D^2} (\tilde A'\tilde B + \tilde A\tilde B') 
\end{split}
\eeq 
\beq
\begin{split}
\frac{\partial \Gtildell'}{\partial r}\biggl|_{r+Yr'} ={}& \frac{2Y}{\tilde D^3} \left[ \tilde C^2(\tilde A\tilde B' + \tilde A'\tilde B + \tilde A\tilde A') - \tilde A^3\tilde B' \right] , \\ 
\frac{\partial \Gtildett'}{\partial r}\biggl|_{r+Yr'} ={}& \frac{2Y}{\tilde D^3} \left[ \tilde C^2(\tilde A\tilde B' + \tilde A'\tilde B + \tilde B\tilde B') - \tilde A'\tilde B^3 \right] , \\ 
\frac{\partial \Gtildelt'}{\partial r}\biggl|_{r+Yr'} ={}& -\frac{Y\tilde C}{\tilde D^3} \Bigl[ (\tilde  C^2-\tilde A\tilde B)(\tilde A' + \tilde B') \\ &- 2\tilde A^2\tilde B' - 2\tilde A'\tilde B^2 \Bigr] , 
\end{split}
\label{B12}
\eeq
and 
\beq
\begin{split}
\frac{\partial\Gtildell'}{\partial r'} &=  - \frac{Y}{\tilde D^2} (\tilde C^2 - \tilde A^2) , \\ 
\frac{\partial\Gtildett'}{\partial r'} &=  - \frac{Y}{\tilde D^2} (\tilde C^2 - \tilde B^2) , \\ 
\frac{\partial\Gtildelt'}{\partial r'} &=  - \frac{Y\tilde C}{\tilde D^2} (\tilde A + \tilde B) .
\end{split}
\eeq
We have introduced $\tilde A'=\partial_y \tilde A$ and $\tilde B'=\partial_y \tilde B$. In (\ref{B11}) and (\ref{B12}), the derivative $\partial/\partial r$ is taken with $r+Yr'$, \ie $R'=\partial_{\q^2}R$, fixed. If $A$, $B$ and $C$ are evaluated within the derivation expansion [Eqs.~(\ref{B7},\ref{B8})],  
\beq
\begin{split}
\tilde A &= \tilde V_A\tilde\w^2 + y +Yr , \\ 
\tilde B &= \tilde A + 2\tilde n_0 \tilde\lambda, \\ 
\tilde C &= \tilde\w ,
\end{split}
\eeq
and
\beq
\tilde A' = \tilde B' = 1+r+Yr' .
\eeq

\section{Solution of the flow equations in the infrared limit}
\label{app_flow_ir} 

In this appendix, we consider the regulator (\ref{regdef}) with~\cite{Litim00}
\beq
r(Y) = \frac{1-Y}{Y} \theta(1-Y) . 
\eeq
We also take 
\beq
c_0^2= \frac{Z_A}{2mV_A} 
\eeq
and note that $c_0$ is $k$-independent in the infrared limit $k\ll k_G$ and equal to the the Goldstone mode velocity $c$ (Secs.~\ref{sec_de} and \ref{sec_nr}). For $Y\leq 1$, one then has
\beq
\begin{split}
\tilde A &= Y+Yr(Y) = 1 , \\ 
\tilde B &= 1 + 2\tilde n_0 \tilde\lambda , \\ 
\tilde D &= 1 + 2\tilde n_0 \tilde\lambda + \tilde\w^2 . 
\end{split}
\eeq
We also observe that the condition $Y\leq 1$ implies 
\beq
|\tilde\w| \leq \frac{Z_C}{Z_A\eps_k} c_0 k \sim k^{2-d} ,
\eeq
where we have anticipated that $Z_C\sim k^{3-d}$ for $d<3$. On the other hand,
\beq
\tilde n_0 \tilde\lambda = (Z_A\eps_k)^{-1} n_0 \lambda \sim k^{1-d} . 
\eeq
We can therefore neglect $\tilde\w^2$ with respect to $\tilde B$, and 
\beq
\tilde D \simeq \tilde B \simeq 2\tilde n_0 \tilde\lambda 
\label{B10}
\eeq
becomes frequency independent. For $d=3$, $|\tilde\w|\lesssim 1/(k\ln k)$ and $\tilde n_0\tilde\lambda \sim 1/(k^2\ln k)$, so that (\ref{B10}) holds. 

We are now in a position to compute the infrared limit of the coefficients $\tilde I_\alpha$ and $\tilde J_{\alpha\beta}$. Since $\eta_A\to 0$, we have
\beq
\begin{split} 
\Itildell &= -4v_d \int_{y,\tilde\w} y^{d/2-1} Y^2r' \frac{\tilde A^2-\tilde\w^2}{\tilde D^2} , \\ 
\Itildett &= -4v_d \int_{y,\tilde\w} y^{d/2-1} Y^2r' \frac{\tilde B^2-\tilde\w^2}{\tilde D^2} .
\end{split}
\eeq
Since $|\tilde\w|,\tilde A\ll \tilde B$, we can neglect $\Itildell$ with respect to $\Itildett$ and approximate
\begin{align}
\Itildett &\simeq -4v_d \int_{y,\tilde\w} y^{d/2-1} Y^2r' \frac{\tilde B^2}{\tilde D^2} \nonumber \\
&= 4v_d \int_{y,\tilde\w} y^{d/2-1} \theta(1-Y) . 
\end{align}
For any function $f(Y)$, 
\begin{multline}
v_d \int_0^\infty dy\, y^{d/2-1} \intinf \frac{d\tilde\w}{2\pi} f(Y) \\ = \tilde V_A^{-1/2} v_{d+1} \int_0^\infty dY\, Y^{(d-1)/2} f(Y) ,
\label{intd}
\end{multline}
so that we finally obtain
\beq
\Itildett \simeq 8 \frac{v_{d+1}}{d+1} \tilde V_A^{-1/2}
\eeq
and
\begin{align}
\dt\tilde n_0 &\simeq -(d+\eta_C) \tilde n_0 + \half \Itildett \nonumber \\ &
\simeq -(d+\eta_C) \tilde n_0 + 4\frac{v_{d+1}}{d+1} \tilde V_A^{-1/2}  \nonumber \\ &
\simeq -(d+\eta_C) \tilde n_0 ,
\label{eq_ir1}
\end{align}
where we have used the fact that the condensate density $n_0$ flows to a finite value when $k\to 0$ (so that the flow of $\tilde n_0$ is determined by the purely dimensional contribution). 

With a similar reasoning, we find
\beq
\begin{split}
\dt\tilde\lambda &\simeq (d-2+\eta_C)\tilde\lambda -\tilde\lambda^2 \Jtildetttt(0) \\ 
&\simeq (d-2+\eta_C)\tilde\lambda + 8 \frac{v_{d+1}}{d+1} \frac{\tilde\lambda^2}{\tilde V_A^{1/2}} , \\ 
\eta_C &\simeq -2\tilde n_0\tilde\lambda^2 \frac{\partial}{\partial\tilde\w} \Jtildettlt(p) \Bigl|_{p=0} \simeq - 8 \frac{v_{d+1}}{d+1} \frac{\tilde\lambda}{\tilde V_A^{1/2}} , \\ 
\dt\tilde V_A &\simeq  (2+2\eta_C) \tilde V_A . 
\end{split} 
\label{eq_ir2}
\eeq

All the integrals involved in the derivation of (\ref{eq_ir1},\ref{eq_ir2}) are $d+1$-dimensional integrals of the type (\ref{intd}). This is a direct manifestation of the relativistic invariance which emerges in the low-energy limit (Sec.~\ref{subsec_analytic}). To compute the infrared limit of the flow equations satisfied by the self-energies, we need to compute the coefficients $\tilde J_{\alpha\beta}(p)$ for finite $p$. The external variable $p$ acts as a low-energy cutoff, so that $\tilde J_{\alpha\beta}(p)$ can be obtained from $\tilde J_{\alpha\beta}(p=0)$ with $k\sim (\p^2+\w^2/c^2)^{1/2}$ (this choice satisfies the relativistic invariance).

\section{High-frequency limit of the two-point vertex} 
\label{app_high_w} 

The normal and anomalous propagators $\Gn$ and $\Gan$ defined in Appendix \ref{app_bog} can be written as
\beq
\begin{split}
\Gn(p) &= \intinf d\w' \frac{\An(\p,\w')}{i\w-\w'} , \\ 
\Gan(p) &= \intinf d\w' \frac{\Aan(\p,\w')}{i\w-\w'} ,
\end{split}
\label{he1}
\eeq
when $i\w\neq 0$. The spectral functions $\An$ and $\Aan$ are defined by
\beq
\begin{split}
\An(\p,t) = \frac{1}{2\pi} \mean{[\hat\psi(\p,t),\hat\psi^\dagger(\p,0)]} , \\
\Aan(\p,t) = \frac{1}{2\pi} \mean{[\hat\psi(\p,t),\hat\psi(-\p,0)]} ,  
\end{split}
\eeq
where $\hat\psi(\p,t)$ and $\hat\psi^\dagger(\p,t)$ are the boson operators in the Heisenberg picture. From the spectral representation (\ref{he1}), we obtain the high-frequency expansion 
\beq
\begin{split}
\Gn(p) &= \frac{1}{i\w} + \frac{X_\p}{(i\w)^2} + \calO(\w^{-3}) , \\ 
\Gan(p) &= \frac{Y_\p}{(i\w)^2} + \calO(\w^{-3}) ,
\end{split}
\eeq
where
\beq
\begin{split}
X_\p ={}& \intinf d\w \w \An(\p,\w) \\ ={}& 2\pi \left[i\dt \An(\p,t)\right]_{t=0} = \mean{\bigl[[\hat\psi(\p),\hat H],\hat\psi^\dagger(\p)\bigr]} , \\ 
Y_\p ={}& \intinf d\w \w \Aan(\p,\w) \\ ={}& 2\pi \left[i\dt \Aan(\p,t)\right]_{t=0} = \mean{\bigl[[\hat\psi(\p),\hat H],\hat\psi(-\p)\bigr]} ,
\end{split}
\label{he2}
\eeq
and $\hat H$ is the quantum Hamiltonian corresponding to the action (\ref{action4}). 
To obtain (\ref{he2}), we have used the equations of motion of the operators $\hat\psi(\p,t)$ and $\hat\psi^\dagger(\p,t)$. A straightforward calculation gives
\beq
\begin{split}
X_\p &= \eps_\p-\mu+2g\bar n , \\
Y_\p &= -g \mean{\hat\psi(\r)^2} ,
\end{split}
\eeq
where $\bar n=\mean{\hat\psi^\dagger(\r)\hat\psi(\r)}$ is the mean boson density. Inverting (\ref{propa}) and considering the high-frequency limit, we obtain
\beq
\begin{split} 
X_\p &= - \lim_{\w\to\infty} \left[\mu-\eps_\p-\Sign(p)\right] , \\
Y_\p &= - \lim_{\w\to\infty} \Sigan(p) ,
\end{split}
\eeq
i.e.
\beq
\begin{split}
\lim_{\w\to\infty} \Sign(p) &= 2g\bar n, \\ 
\lim_{\w\to\infty} \Sigan(p) &= g \mean{\hat\psi(\r)^2} .
\end{split}
\eeq
From (\ref{sigrel}) and (\ref{deltadef}), we finally deduce 
\beq
\begin{split}
\bar\Delta_A^\infty &= \lim_{\w\to\infty} \bar\Delta_A(p) = -\mu-g \mean{\hat\psi(\r)^2} + 2g\bar n, \\ 
\bar\Gamma_B^\infty &= \lim_{\w\to\infty} \bar\Gamma_B(p) = g \frac{\mean{\hat\psi(\r)^2}}{n_0} , \\
\bar\Delta_C^\infty &= \lim_{\w\to\infty} \bar\Delta_C(p) = 0 . 
\end{split}
\label{he3}
\eeq
From these limiting values, we can obtain the ``pairing'' amplitude $\mean{\hat\psi(\r)^2}$ and the mean boson density $\bar n$. In the weak-coupling limit, $\bar n\simeq n_0 \simeq \mean{\hat\psi(\r)^2}$ and $n_0\simeq \mu/g$ does not differ much from the Bogoliubov result, so that we expect $\bar\Delta_A^\infty\ll g\bar n$, $\bar\Gamma_B^\infty\simeq g$ and $\bar\Delta_C^\infty=0$.

Since $\lim_{\w\to\infty} \bar J_{\alpha\beta}(p)=0$, the flow equations (\ref{flow5},\ref{flow3}) yield
\beq
\begin{split} 
\dt \bar\Delta_A^\infty &= \lambda\left( \Ibarll-\Ibartt \right) , \\
\dt \bar\Gamma_B^\infty &= \frac{\lambda}{2n_0} \left( \Ibartt-\Ibarll \right) , \\
\dt \bar\Delta_C^\infty &= 0 . 
\end{split}
\label{he4}
\eeq
These equations are not exact as they involve $\lambda\sim\Gamma^{(4)}(0,0,0,0)$ rather than the high-frequency limit $\lim_{\wnu\to\infty}\Gamma^{(4)}(p,-p,q,-q)$ (with $p=(\p,i\wnu)$) of the four-point vertex. Nevertheless, the numerical results of Sec.~\ref{sec_nr} are in good agreement with the asymptotic values (\ref{he3}). Note that contrary to (\ref{he4}), the BMW equations would be correct in the high-frequency limit.

%\bibliography{/users/lptl/dupuis/publi/BIB/bosons.bib,/users/lptl/dupuis/publi/BIB/nprg.bib,/users/lptl/dupuis/publi/BIB/divers.bib,/users/lptl/dupuis/publi/BIB/book.bib}

\begin{thebibliography}{99}
\expandafter\ifx\csname natexlab\endcsname\relax\def\natexlab#1{#1}\fi
\expandafter\ifx\csname bibnamefont\endcsname\relax
  \def\bibnamefont#1{#1}\fi
\expandafter\ifx\csname bibfnamefont\endcsname\relax
  \def\bibfnamefont#1{#1}\fi
\expandafter\ifx\csname citenamefont\endcsname\relax
  \def\citenamefont#1{#1}\fi
\expandafter\ifx\csname url\endcsname\relax
  \def\url#1{\texttt{#1}}\fi
\expandafter\ifx\csname urlprefix\endcsname\relax\def\urlprefix{URL }\fi
\providecommand{\bibinfo}[2]{#2}
\providecommand{\eprint}[2][]{\url{#2}}

\bibitem[{\citenamefont{Bogoliubov}(1947)}]{Bogoliubov47}
\bibinfo{author}{\bibfnamefont{N.~N.} \bibnamefont{Bogoliubov}},
  \bibinfo{journal}{J. Phys. USSR} \textbf{\bibinfo{volume}{11}},
  \bibinfo{pages}{23} (\bibinfo{year}{1947}).

\bibitem[{not({\natexlab{a}})}]{note4}
\bibinfo{note}{For a review, see H. Shi and A. Griffin, Phys. Rep. {\bf 304}, 1
  (1998); J.~O. Andersen, Rev. Mod. Phys. {\bf 76}, 599 (2004).}

\bibitem[{\citenamefont{Beliaev}(1958{\natexlab{a}})}]{Beliaev58a}
\bibinfo{author}{\bibfnamefont{S.~T.} \bibnamefont{Beliaev}},
  \bibinfo{journal}{Sov. Phys. JETP} \textbf{\bibinfo{volume}{7}},
  \bibinfo{pages}{289} (\bibinfo{year}{1958}{\natexlab{a}}).

\bibitem[{\citenamefont{Beliaev}(1958{\natexlab{b}})}]{Beliaev58b}
\bibinfo{author}{\bibfnamefont{S.~T.} \bibnamefont{Beliaev}},
  \bibinfo{journal}{Sov. Phys. JETP} \textbf{\bibinfo{volume}{7}},
  \bibinfo{pages}{299} (\bibinfo{year}{1958}{\natexlab{b}}).

\bibitem[{\citenamefont{Hugenholtz and Pines}(1959)}]{Hugenholtz59}
\bibinfo{author}{\bibfnamefont{N.}~\bibnamefont{Hugenholtz}} \bibnamefont{and}
  \bibinfo{author}{\bibfnamefont{D.}~\bibnamefont{Pines}},
  \bibinfo{journal}{Phys. Rev.} \textbf{\bibinfo{volume}{116}},
  \bibinfo{pages}{489} (\bibinfo{year}{1959}).

\bibitem[{\citenamefont{Gavoret and Nozi\`eres}(1964)}]{Gavoret64}
\bibinfo{author}{\bibfnamefont{J.}~\bibnamefont{Gavoret}} \bibnamefont{and}
  \bibinfo{author}{\bibfnamefont{P.}~\bibnamefont{Nozi\`eres}},
  \bibinfo{journal}{Ann. Phys. (N.Y.)} \textbf{\bibinfo{volume}{28}},
  \bibinfo{pages}{349} (\bibinfo{year}{1964}).

\bibitem[{\citenamefont{Patasinskij and Pokrovskij}(1973)}]{Patasinskij73}
\bibinfo{author}{\bibfnamefont{A.~Z.} \bibnamefont{Patasinskij}}
  \bibnamefont{and} \bibinfo{author}{\bibfnamefont{V.~L.}
  \bibnamefont{Pokrovskij}}, \bibinfo{journal}{Sov. Phys. JETP}
  \textbf{\bibinfo{volume}{37}}, \bibinfo{pages}{733} (\bibinfo{year}{1973}).

\bibitem[{\citenamefont{Nepomnyashchii and
  Nepomnyashchii}(1975)}]{Nepomnyashchii75}
\bibinfo{author}{\bibfnamefont{A.~A.} \bibnamefont{Nepomnyashchii}}
  \bibnamefont{and} \bibinfo{author}{\bibfnamefont{Y.~A.}
  \bibnamefont{Nepomnyashchii}}, \bibinfo{journal}{JETP Lett.}
  \textbf{\bibinfo{volume}{21}}, \bibinfo{pages}{1} (\bibinfo{year}{1975}).

\bibitem[{\citenamefont{Nepomnyashchii and
  Nepomnyashchii}(1978)}]{Nepomnyashchii78}
\bibinfo{author}{\bibfnamefont{Y.~A.} \bibnamefont{Nepomnyashchii}}
  \bibnamefont{and} \bibinfo{author}{\bibfnamefont{A.~A.}
  \bibnamefont{Nepomnyashchii}}, \bibinfo{journal}{Sov. Phys. JETP}
  \textbf{\bibinfo{volume}{48}}, \bibinfo{pages}{493} (\bibinfo{year}{1978}).

\bibitem[{\citenamefont{Nepomnyashchii}(1983)}]{Nepomnyashchii83}
\bibinfo{author}{\bibfnamefont{Y.~A.} \bibnamefont{Nepomnyashchii}},
  \bibinfo{journal}{Sov. Phys. JETP} \textbf{\bibinfo{volume}{58}},
  \bibinfo{pages}{722} (\bibinfo{year}{1983}).

\bibitem[{\citenamefont{Popov}(1983)}]{Popov_book_2}
\bibinfo{author}{\bibfnamefont{V.~N.} \bibnamefont{Popov}},
  \emph{\bibinfo{title}{Functional Integrals in Quantum Field Theory and
  Statistical Physics}} (\bibinfo{publisher}{Reidel},
  \bibinfo{address}{Dordrecht, Holland}, \bibinfo{year}{1983}).

\bibitem[{\citenamefont{Popov and Seredniakov}(1979)}]{Popov79}
\bibinfo{author}{\bibfnamefont{V.~N.} \bibnamefont{Popov}} \bibnamefont{and}
  \bibinfo{author}{\bibfnamefont{A.~V.} \bibnamefont{Seredniakov}},
  \bibinfo{journal}{Sov. Phys. JETP} \textbf{\bibinfo{volume}{50}},
  \bibinfo{pages}{193} (\bibinfo{year}{1979}).

\bibitem[{\citenamefont{Giorgini et~al.}(1992)\citenamefont{Giorgini,
  Pitaevskii, and Stringari}}]{Giorgini92}
\bibinfo{author}{\bibfnamefont{S.}~\bibnamefont{Giorgini}},
  \bibinfo{author}{\bibfnamefont{L.}~\bibnamefont{Pitaevskii}},
  \bibnamefont{and}
  \bibinfo{author}{\bibfnamefont{S.}~\bibnamefont{Stringari}},
  \bibinfo{journal}{Phys. Rev. B} \textbf{\bibinfo{volume}{46}},
  \bibinfo{pages}{6374} (\bibinfo{year}{1992}).

\bibitem[{\citenamefont{Castellani et~al.}(1997)\citenamefont{Castellani,
  Castro, Pistolesi, and Strinati}}]{Castellani97}
\bibinfo{author}{\bibfnamefont{C.}~\bibnamefont{Castellani}},
  \bibinfo{author}{\bibfnamefont{C.} \bibnamefont{Di Castro}},
  \bibinfo{author}{\bibfnamefont{F.}~\bibnamefont{Pistolesi}},
  \bibnamefont{and} \bibinfo{author}{\bibfnamefont{G.~C.}
  \bibnamefont{Strinati}}, \bibinfo{journal}{Phys. Rev. Lett.}
  \textbf{\bibinfo{volume}{78}}, \bibinfo{pages}{1612} (\bibinfo{year}{1997}).

\bibitem[{\citenamefont{Pistolesi et~al.}(2004)\citenamefont{Pistolesi,
  Castellani, Castro, and Strinati}}]{Pistolesi04}
\bibinfo{author}{\bibfnamefont{F.}~\bibnamefont{Pistolesi}},
  \bibinfo{author}{\bibfnamefont{C.}~\bibnamefont{Castellani}},
  \bibinfo{author}{\bibfnamefont{C.} \bibnamefont{Di Castro}},
  \bibnamefont{and} \bibinfo{author}{\bibfnamefont{G.~C.}
  \bibnamefont{Strinati}}, \bibinfo{journal}{Phys. Rev. B}
  \textbf{\bibinfo{volume}{69}}, \bibinfo{pages}{024513}
  (\bibinfo{year}{2004}).

\bibitem[{\citenamefont{Wetterich}(2008)}]{Wetterich08}
\bibinfo{author}{\bibfnamefont{C.}~\bibnamefont{Wetterich}},
  \bibinfo{journal}{Phys. Rev. B} \textbf{\bibinfo{volume}{77}},
  \bibinfo{eid}{064504} (\bibinfo{year}{2008}).

\bibitem[{\citenamefont{Dupuis and Sengupta}(2007)}]{Dupuis07}
\bibinfo{author}{\bibfnamefont{N.}~\bibnamefont{Dupuis}} \bibnamefont{and}
  \bibinfo{author}{\bibfnamefont{K.}~\bibnamefont{Sengupta}},
  \bibinfo{journal}{Europhys. Lett.} \textbf{\bibinfo{volume}{80}},
  \bibinfo{pages}{50007} (\bibinfo{year}{2007}).

\bibitem[{\citenamefont{Sinner et~al.}(2009)\citenamefont{Sinner, Hasselmann,
  and Kopietz}}]{Sinner09}
\bibinfo{author}{\bibfnamefont{A.}~\bibnamefont{Sinner}},
  \bibinfo{author}{\bibfnamefont{N.}~\bibnamefont{Hasselmann}},
  \bibnamefont{and} \bibinfo{author}{\bibfnamefont{P.}~\bibnamefont{Kopietz}},
  \bibinfo{journal}{Phys. Rev. Lett.} \textbf{\bibinfo{volume}{102}},
  \bibinfo{pages}{120601} (\bibinfo{year}{2009}).

\bibitem[{\citenamefont{Dupuis}(2009)}]{Dupuis09}
\bibinfo{author}{\bibfnamefont{N.}~\bibnamefont{Dupuis}},
  \bibinfo{journal}{Phys. Rev. Lett.} \textbf{\bibinfo{volume}{102}},
  \bibinfo{pages}{190401} (\bibinfo{year}{2009}).

\bibitem{Andersen99} J.~O. Andersen and M. Strickland, Phys. Rev. A {\bf 60}, 1442 (1999). 

\bibitem[{\citenamefont{Floerchinger and Wetterich}(2008)}]{Floerchinger08}
\bibinfo{author}{\bibfnamefont{S.}~\bibnamefont{Floerchinger}}
  \bibnamefont{and}
  \bibinfo{author}{\bibfnamefont{C.}~\bibnamefont{Wetterich}},
  \bibinfo{journal}{Phys. Rev. A} \textbf{\bibinfo{volume}{77}},
  \bibinfo{eid}{053603} (\bibinfo{year}{2008}).

\bibitem[{\citenamefont{Floerchinger and
  Wetterich}(2009{\natexlab{a}})}]{Floerchinger09a}
\bibinfo{author}{\bibfnamefont{S.}~\bibnamefont{Floerchinger}}
  \bibnamefont{and}
  \bibinfo{author}{\bibfnamefont{C.}~\bibnamefont{Wetterich}},
  \bibinfo{journal}{Phys. Rev. A} \textbf{\bibinfo{volume}{79}},
  \bibinfo{eid}{013601} (\bibinfo{year}{2009}{\natexlab{a}}).

\bibitem[{\citenamefont{Floerchinger and
  Wetterich}(2009{\natexlab{b}})}]{Floerchinger09b}
\bibinfo{author}{\bibfnamefont{S.}~\bibnamefont{Floerchinger}}
  \bibnamefont{and}
  \bibinfo{author}{\bibfnamefont{C.}~\bibnamefont{Wetterich}},
  \bibinfo{journal}{Phys. Rev. A} \textbf{\bibinfo{volume}{79}},
  \bibinfo{eid}{063602} (\bibinfo{year}{2009}{\natexlab{b}}).

\bibitem[{\citenamefont{Eichler et~al.}()\citenamefont{Eichler, Hasselmann, and
  Kopietz}}]{Eichler09}
\bibinfo{author}{\bibfnamefont{C.}~\bibnamefont{Eichler}},
  \bibinfo{author}{\bibfnamefont{N.}~\bibnamefont{Hasselmann}},
  \bibnamefont{and} \bibinfo{author}{\bibfnamefont{P.}~\bibnamefont{Kopietz}},
  \bibinfo{note}{arXiv:0906.0868}.

\bibitem[{\citenamefont{Blaizot et~al.}(2006)\citenamefont{Blaizot,
  M\'endez-Galain, and Wschebor}}]{Blaizot06}
\bibinfo{author}{\bibfnamefont{J.-P.} \bibnamefont{Blaizot}},
  \bibinfo{author}{\bibfnamefont{R.}~\bibnamefont{M\'endez-Galain}},
  \bibnamefont{and} \bibinfo{author}{\bibfnamefont{N.}~\bibnamefont{Wschebor}},
  \bibinfo{journal}{Phys. Lett. B} \textbf{\bibinfo{volume}{632}},
  \bibinfo{pages}{571} (\bibinfo{year}{2006}).

\bibitem{Benitez09}
F. Benitez, J.~P. Blaizot, H. Chat\'e, B. Delamotte, R. M\'endez-Galain and N. Wschebor, Phys. Rev. E {\bf 80}, 030103(R) (2009).

\bibitem[{\citenamefont{Weichman}(1988)}]{Weichman88}
\bibinfo{author}{\bibfnamefont{P.~B.} \bibnamefont{Weichman}},
  \bibinfo{journal}{Phys. Rev. B} \textbf{\bibinfo{volume}{38}},
  \bibinfo{pages}{8739} (\bibinfo{year}{1988}).

\bibitem{Zinn_book} See, for instance, J. Zinn-Justin, {\it Quantum Field Theory and Critical Phenomena}, chapter 27 (Third Edition, Clarendon Press, Oxford, 1996). 

\bibitem[{\citenamefont{Fisher et~al.}(1973)\citenamefont{Fisher, Barber, and
  Jasnow}}]{Fisher73}
\bibinfo{author}{\bibfnamefont{M.~E.} \bibnamefont{Fisher}},
  \bibinfo{author}{\bibfnamefont{M.~N.} \bibnamefont{Barber}},
  \bibnamefont{and} \bibinfo{author}{\bibfnamefont{D.}~\bibnamefont{Jasnow}},
  \bibinfo{journal}{Phys. Rev. A} \textbf{\bibinfo{volume}{8}},
  \bibinfo{pages}{1111} (\bibinfo{year}{1973}).

\bibitem[{\citenamefont{Anishetty et~al.}(1999)\citenamefont{Anishetty, Basu,
  Dass, and Sharatchandra}}]{Anishetty99}
\bibinfo{author}{\bibfnamefont{R.}~\bibnamefont{Anishetty}},
  \bibinfo{author}{\bibfnamefont{R.}~\bibnamefont{Basu}},
  \bibinfo{author}{\bibfnamefont{N.~D.} \bibnamefont{Hari Dass}},
  \bibnamefont{and} \bibinfo{author}{\bibfnamefont{H.~S.}
  \bibnamefont{Sharatchandra}}, \bibinfo{journal}{Int. J. Mod. Phys. A}
  \textbf{\bibinfo{volume}{14}}, \bibinfo{pages}{3467} (\bibinfo{year}{1999}).

\bibitem[{\citenamefont{Berges et~al.}(2002)\citenamefont{Berges, Tetradis, and
  Wetterich}}]{Berges02}
\bibinfo{author}{\bibfnamefont{J.}~\bibnamefont{Berges}},
  \bibinfo{author}{\bibfnamefont{N.}~\bibnamefont{Tetradis}}, \bibnamefont{and}
  \bibinfo{author}{\bibfnamefont{C.}~\bibnamefont{Wetterich}},
  \bibinfo{journal}{Phys. Rep.} \textbf{\bibinfo{volume}{363}},
  \bibinfo{pages}{223} (\bibinfo{year}{2002}).

\bibitem[{\citenamefont{Zwerger}(2004)}]{Zwerger04}
\bibinfo{author}{\bibfnamefont{W.}~\bibnamefont{Zwerger}},
  \bibinfo{journal}{Phys. Rev. Lett.} \textbf{\bibinfo{volume}{92}},
  \bibinfo{pages}{027203} (\bibinfo{year}{2004}).

\bibitem{note2} i.e. the invariance in the transformation $\psi(\r,\tau)\to e^{i\alpha} \psi(\r,\tau)$, $\psi^*(\r,\tau)\to e^{-i\alpha} \psi^*(\r,\tau)$. This transformation corresponds to a rotation of angle $\alpha$ of the two-component real field $(\psi_1,\psi_2)$.

\bibitem[{\citenamefont{Abrikosov et~al.}(1975)\citenamefont{Abrikosov,
  Gor'kov, and Dzyaloshinski}}]{AGD_book}
\bibinfo{author}{\bibfnamefont{A.~A.} \bibnamefont{Abrikosov}},
  \bibinfo{author}{\bibfnamefont{L.~P.} \bibnamefont{Gor'kov}},
  \bibnamefont{and} \bibinfo{author}{\bibfnamefont{I.~E.}
  \bibnamefont{Dzyaloshinski}}, \emph{\bibinfo{title}{Methods of Quantum Field
  Theory in Statistical Physics}} (\bibinfo{publisher}{Dover},
  \bibinfo{year}{1975}).

\bibitem[{\citenamefont{Fetter and Walecka}(2003)}]{Fetter_book}
\bibinfo{author}{\bibfnamefont{A.~L.} \bibnamefont{Fetter}} \bibnamefont{and}
  \bibinfo{author}{\bibfnamefont{J.~D.} \bibnamefont{Walecka}},
  \emph{\bibinfo{title}{Quantum Theory of Many-Particle Systems}}
  (\bibinfo{publisher}{Dover}, \bibinfo{year}{2003}).

\bibitem[{not({\natexlab{b}})}]{note8}
\bibinfo{note}{Note that in the weak coupling limit where $n_0\simeq \mu/g$,
  $k_h=\sqrt{2gmn_0}\simeq \sqrt{2m\mu}$ is roughly independent of $g$ if the
  chemical potential (rather than the mean boson density $\bar n\simeq n_0$) is
  fixed. We then find $k_G\sim g^{1/(3-d)}$ (for $d<3$) as in the
  $(d+1)$-dimensional classical $O(2)$ model.}

\bibitem[{not({\natexlab{c}})}]{note1}
\bibinfo{note}{Alternatively, one could write
  $\Gamma_{ij}^{(2)}=\frac{\phi_i\phi_j}{2n} \Gamma_{\rm ll} +
  \left(\delta_{ij} - \frac{\phi_i\phi_j}{2n}\right) \Gamma_{\rm tt} +
  \eps_{ij} \Gamma_{\rm lt}$ in terms of longitudinal (l) and transverse (t)
  fluctuations, with $\Gamma_{\rm ll}=\Gamma_A+2n \Gamma_B$, $\Gamma_{\rm
  tt}=\Gamma_A$ and $\Gamma_{\rm lt}=\Gamma_C$.}

\bibitem[{\citenamefont{Huang and Klein}(1964)}]{Huang64}
\bibinfo{author}{\bibfnamefont{K.}~\bibnamefont{Huang}} \bibnamefont{and}
  \bibinfo{author}{\bibfnamefont{A.}~\bibnamefont{Klein}},
  \bibinfo{journal}{Ann. Phys. (N.Y.)} \textbf{\bibinfo{volume}{30}},
  \bibinfo{pages}{203} (\bibinfo{year}{1964}).

\bibitem[{not({\natexlab{d}})}]{note7}
\bibinfo{note}{Because of the infrared regulator $R$, the propagator $\bar
  G(p)=-\bar\Gamma^{(2)-1}(p)-R(p)$ entering the flow equations has a gap $ck$
  [Eqs.~(\ref{B4})] (while the propagator $-\bar\Gamma^{(2)-1}(p)$ is gapless
  in agreement with Goldstone theorem). This property ensures that the
  $n$-point vertex $\bar\Gamma^{(n)}(p_1,\cdots,p_n)$ is a regular function of
  its arguments for $|\p_i|,|\w_i|/c\ll k$.}

\bibitem[{\citenamefont{Chaikin and Lubensky}(1995)}]{Chaikin_book}
\bibinfo{author}{\bibfnamefont{P.~M.} \bibnamefont{Chaikin}} \bibnamefont{and}
  \bibinfo{author}{\bibfnamefont{T.~C.} \bibnamefont{Lubensky}},
  \emph{\bibinfo{title}{Principles of Condensed Matter Physics}}
  (\bibinfo{publisher}{Cambridge University Press}, \bibinfo{year}{1995}).

\bibitem[{\citenamefont{Benitez et~al.}(2008)\citenamefont{Benitez,
  M\'{e}ndez-Galain, and Wschebor}}]{Benitez08}
\bibinfo{author}{\bibfnamefont{F.}~\bibnamefont{Benitez}},
  \bibinfo{author}{\bibfnamefont{R.}~\bibnamefont{M\'{e}ndez-Galain}},
  \bibnamefont{and} \bibinfo{author}{\bibfnamefont{N.}~\bibnamefont{Wschebor}},
  \bibinfo{journal}{Phys. Rev. B} \textbf{\bibinfo{volume}{77}},
  \bibinfo{pages}{024431} (\bibinfo{year}{2008}).

\bibitem[{\citenamefont{Guerra et~al.}(2007)\citenamefont{Guerra,
  M\'endez-Galain, and Wschebor}}]{Guerra07}
\bibinfo{author}{\bibfnamefont{D.}~\bibnamefont{Guerra}},
  \bibinfo{author}{\bibfnamefont{R.}~\bibnamefont{M\'endez-Galain}},
  \bibnamefont{and} \bibinfo{author}{\bibfnamefont{N.}~\bibnamefont{Wschebor}},
  \bibinfo{journal}{Eur. Phys. J. B} \textbf{\bibinfo{volume}{59}},
  \bibinfo{pages}{357} (\bibinfo{year}{2007}).

\bibitem[{\citenamefont{Wetterich}(1993)}]{Wetterich93}
\bibinfo{author}{\bibfnamefont{C.}~\bibnamefont{Wetterich}},
  \bibinfo{journal}{Phys. Lett. B} \textbf{\bibinfo{volume}{301}},
  \bibinfo{pages}{90} (\bibinfo{year}{1993}).

\bibitem[{\citenamefont{Sinner et~al.}(2008)\citenamefont{Sinner, Hasselmann,
  and Kopietz}}]{Sinner08}
\bibinfo{author}{\bibfnamefont{A.}~\bibnamefont{Sinner}},
  \bibinfo{author}{\bibfnamefont{N.}~\bibnamefont{Hasselmann}},
  \bibnamefont{and} \bibinfo{author}{\bibfnamefont{P.}~\bibnamefont{Kopietz}},
  \bibinfo{journal}{J. Phys.: Cond. Matt.} \textbf{\bibinfo{volume}{20}},
  \bibinfo{pages}{075208} (\bibinfo{year}{2008}).

\bibitem[{\citenamefont{Dupuis and Sengupta}(2008)}]{Dupuis08}
\bibinfo{author}{\bibfnamefont{N.}~\bibnamefont{Dupuis}} \bibnamefont{and}
  \bibinfo{author}{\bibfnamefont{K.}~\bibnamefont{Sengupta}},
  \bibinfo{journal}{Eur. Phys. J. B} \textbf{\bibinfo{volume}{66}},
  \bibinfo{pages}{271} (\bibinfo{year}{2008}).

\bibitem[{not({\natexlab{e}})}]{note3}
\bibinfo{note}{$\lambda$ also shows a weak maximum at $k\sim k_h$.}

\bibitem[{\citenamefont{Vidberg and Serene}(1977)}]{Vidberg77}
\bibinfo{author}{\bibfnamefont{H.~J.} \bibnamefont{Vidberg}} \bibnamefont{and}
  \bibinfo{author}{\bibfnamefont{J.~W.} \bibnamefont{Serene}},
  \bibinfo{journal}{J. Low Temp. Phys} \textbf{\bibinfo{volume}{29}},
  \bibinfo{pages}{179} (\bibinfo{year}{1977}).

\bibitem{Lifshitz_stat_phys_II} See, for instance, E.~M. Lifshitz and L.~P. Pitaevskii, {\it Statistical Physics II}  (Pergamon, Oxford, 1980).

\bibitem[{\citenamefont{Kreisel et~al.}(2008)\citenamefont{Kreisel, Sauli,
  Hasselmann, and Kopietz}}]{Kreisel08}
\bibinfo{author}{\bibfnamefont{A.}~\bibnamefont{Kreisel}},
  \bibinfo{author}{\bibfnamefont{F.}~\bibnamefont{Sauli}},
  \bibinfo{author}{\bibfnamefont{N.}~\bibnamefont{Hasselmann}},
  \bibnamefont{and} \bibinfo{author}{\bibfnamefont{P.}~\bibnamefont{Kopietz}},
  \bibinfo{journal}{Phys. Rev. B} \textbf{\bibinfo{volume}{78}},
  \bibinfo{eid}{035127} (\bibinfo{year}{2008}).

\bibitem[{\citenamefont{Chung and Bhattacherjee}()}]{Chung08}
\bibinfo{author}{\bibfnamefont{M.-C.} \bibnamefont{Chung}} \bibnamefont{and}
  \bibinfo{author}{\bibfnamefont{A.~B.} \bibnamefont{Bhattacherjee}},
  \bibinfo{note}{arXiv:0809.3632}.

\bibitem[{not({\natexlab{f}})}]{note5}
\bibinfo{note}{When the correlation function involves two operators with
  opposite signatures under time reversal, the spectral function is given by
  $i$ times the real part of the retarded correlation function. All spectral
  functions in (\ref{nr5}) satisfy $G_\alpha(\p,z) = \intinf d\w
  \frac{A_\alpha(\p,\w)}{z-\w}$, with $z$ an arbitrary complex frequency.}

\bibitem[{\citenamefont{Khodas et~al.}()\citenamefont{Khodas, Kamenev, and
  Glazman}}]{Khodas08}
\bibinfo{author}{\bibfnamefont{M.}~\bibnamefont{Khodas}},
  \bibinfo{author}{\bibfnamefont{A.}~\bibnamefont{Kamenev}}, \bibnamefont{and}
  \bibinfo{author}{\bibfnamefont{L.~I.} \bibnamefont{Glazman}},
  \bibinfo{note}{arXiv:0710.2910, arXiv:0807.2393.}

\bibitem[{\citenamefont{Lieb and Liniger}(1963)}]{Lieb63a}
\bibinfo{author}{\bibfnamefont{E.~H.} \bibnamefont{Lieb}} \bibnamefont{and}
  \bibinfo{author}{\bibfnamefont{W.}~\bibnamefont{Liniger}},
  \bibinfo{journal}{Phys. Rev.} \textbf{\bibinfo{volume}{130}},
  \bibinfo{pages}{1605} (\bibinfo{year}{1963}).

\bibitem[{\citenamefont{Lieb}(1963)}]{Lieb63b}
\bibinfo{author}{\bibfnamefont{E.~H.} \bibnamefont{Lieb}},
  \bibinfo{journal}{Phys. Rev.} \textbf{\bibinfo{volume}{130}},
  \bibinfo{pages}{1616} (\bibinfo{year}{1963}).

\bibitem[{\citenamefont{Litim}(2000)}]{Litim00}
\bibinfo{author}{\bibfnamefont{D.}~\bibnamefont{Litim}},
  \bibinfo{journal}{Phys. Lett. B} \textbf{\bibinfo{volume}{486}},
  \bibinfo{pages}{92} (\bibinfo{year}{2000}).

\end{thebibliography}
%\bibliographystyle{apsrev}

\end{document}